\documentclass[preprint,12pt]{elsarticle}

\usepackage{amsmath,amssymb,amsthm}
\usepackage{graphicx}
\usepackage{caption}
\usepackage{subcaption}
\usepackage{wrapfig}
\usepackage{dcolumn}
\usepackage{bm}
\usepackage{xcolor}

\definecolor{tab_blue}{HTML}{1F77B4}
\definecolor{tab_orange}{HTML}{FF7F0E}
\definecolor{tab_green}{HTML}{2CA02C}
\definecolor{tab_gray}{HTML}{7F7F7F}

\graphicspath{{figures/}}

\usepackage{hyperref}
\usepackage{inconsolata}

\newcommand{\dee}{\mathrm{d}}

\newcommand{\mb}{\mathbf}
\newcommand{\given}{\:\vert\:}

\newcommand{\Dt}{\Delta_t}
\newcommand{\Deff}{D_{\mathrm{eff}}}
\newcommand{\GP}{\mathcal{GP}}
\newcommand{\order}{\mathcal{O}}

\DeclareMathOperator*{\argmax}{arg\,max}

\journal{Journal of Computational Physics}

\begin{document}

\title{Low-rank statistical finite elements for scalable model-data synthesis}

\author[uwa,cam]{Connor Duffin\corref{cor}}
\ead{cpd32@cam.ac.uk}
\author[uwa]{Edward Cripps}
\author[uwa,cplx]{Thomas Stemler}
\author[cam,turing]{Mark Girolami}

\affiliation[uwa]{
  organization={Department of Mathematics and Statistics},
  addressline={The University of Western Australia},
  city={Crawley},
  postcode={6009},
  state={WA},
  country={Australia}
}

\affiliation[cam]{
  organization={Department of Engineering},
  addressline={University of Cambridge},
  city={Cambridge},
  postcode={CB2 1PZ},
  country={UK}
}

\affiliation[cplx]{
  organization={Complex Systems Group},
  addressline={The University of Western Australia},
  city={Crawley},
  postcode={6009},
  state={WA},
  country={Australia}
}

\affiliation[turing]{
  organization={Lloyd's Register Programme for Data-Centric Engineering},
  addressline={The Alan Turing Institute},
  city={London},
  postcode={NW1 2DB},
  country={UK}
}

\cortext[cor]{Corresponding author at: Department of Engineering, Trumpington
  Street, Cambridge, CB2 1PZ, UK.}

\date{\today}
\begin{abstract}
Statistical learning additions to physically derived mathematical models are
gaining traction in the literature. A recent approach has been to augment the
underlying physics of the governing equations with data driven Bayesian
statistical methodology. Coined statFEM, the method acknowledges \textit{a
priori} model misspecification, by embedding stochastic forcing within the
governing equations. Upon receipt of additional data, the posterior distribution
of the discretised finite element solution is updated using classical Bayesian
filtering techniques. The resultant posterior jointly quantifies uncertainty
associated with the ubiquitous problem of model misspecification and the data
intended to represent the true process of interest. Despite this appeal,
computational scalability is a challenge to statFEM's application to
high-dimensional problems typically experienced in physical and industrial
contexts. This article overcomes this hurdle by embedding a low-rank
approximation of the underlying dense covariance matrix, obtained from the
leading order modes of the full-rank alternative. Demonstrated on a series of
reaction-diffusion problems of increasing dimension, using experimental and
simulated data, the method reconstructs the sparsely observed data-generating
processes with minimal loss of information, in both the posterior mean and
variance, paving the way for further integration of physical and probabilistic
approaches to complex systems.
\end{abstract}


\maketitle

% \linenumbers\relax

\section{Introduction}

The statistical finite element method
(statFEM)~\cite{girolamiStatisticalFiniteElement2021} provides a novel
statistical construction of the finite element method (FEM) to reconcile
imperfect models with observational data. Potential model misspecification, as
represented by a Gaussian process ($\GP$), is addressed through sequentially
updating the FEM coefficients upon the arrival of observed data, within a
filtering context, to compute a physics-informed posterior distribution.
In~\cite{duffinStatisticalFiniteElements2021}, it is seen that for 1D nonlinear
partial differential equations (PDEs), statFEM is able to give an accurate
approximation to the underlying data generating process, with uncertainty
quantification (UQ), synthesising physics and data to give an interpretable posterior
distribution. Utilising such an approach allows for the application of simpler
computational models, correcting for their possible deficiencies with data.
This approach is also shown to be beneficial in small data regimes.

Recent advances in both the computational physics and machine learning
communities have seen additional approaches proposed for combining physics and
data. In~\cite{rudyDatadrivenDiscoveryPartial2017}, a sparse regression approach
is taken, to learn the underlying PDE from spatio-temporal measurements inside
the problem domain, building on the similar work
of~\cite{bruntonDiscoveringGoverningEquations2016}, to address system
identification from a library of candidate functions. Neural networks can also
be used, for both data-driven solution and system identification, and also for
estimating optimal discretisations from
data~\cite{raissiPhysicsinformedNeuralNetworks2019,bar-sinaiLearningDatadrivenDiscretizations2019}.
Perhaps the most similar to statFEM is~\cite{raissiHiddenPhysicsModels2018},
where a $\GP$ prior is placed over the PDE variable, making the covariance
kernel encode all \textit{a priori} physical assumptions. This allows for system
identification through transforming PDE coefficients into $\GP$ kernel
hyperparameters, the estimation of which is
well-studied~\cite{williamsGaussianProcessesMachine2006}.

StatFEM, instead, places a $\GP$ prior over the forcing of the PDE and focuses
on updating solutions, combining methodology from Bayesian
inversion~\cite{stuartInverseProblemsBayesian2010} and data
assimilation~\cite{lawDataAssimilationMathematical2015}. As in Bayesian
inversion, statFEM uses a Gaussian prior placed over a function inside of the
governing PDE. Though instead of estimating the posterior distribution of this
unknown function, statFEM proceeds to sequentially update the discretised PDE
solution, as in data assimilation. However the model is not treated as a
black-box function, as we embed the uncertainty inside of the PDE formulation,
and discretise the stochastic dynamics.

As presented, applying the statFEM methodology to high-dimensional systems is
challenging. For the successful adoption of statFEM a scalable formulation for
high-dimensional model-data synthesis is required, and this is the focus of this
paper, for, in particular, nonlinear, time-dependent problems. We build on the
extended Kalman filter (ExKF,
\cite{lawDataAssimilationMathematical2015}) method
of~\cite{duffinStatisticalFiniteElements2021}, and study a scalable version of
the statFEM methodology in the context of reaction-diffusion (RD) systems. To
scale the method, we make a low-rank approximation to the posterior covariance
matrix, and embed this within the ExKF;
see~\cite{verlaanTidalFlowForecasting1997,gillijnsReducedRankTransform2006,rozierReducedOrderKalmanFilter2007,lawEvaluatingDataAssimilation2012},
for similar approaches.

Common examples of low-rank filters are the particle filter
\cite{doucetSequentialMonteCarlo2000} and the ensemble Kalman filter (EnKF)
\cite{evensenEnsembleKalmanFilter2003}. However for high-dimensional systems the
particle filter is known to suffer from particle collapse
\cite{bengtssonCurseofdimensionalityRevisitedCollapse2008}, and the ensemble
Kalman filter, whilst accurate for the mean, can fail to provide accurate
UQ \cite{lawEvaluatingDataAssimilation2012}. Our
interest in the low-rank ExKF (LR-ExKF) is motivated by additional results in
\cite{lawEvaluatingDataAssimilation2012}, which show that a low-rank extended
Kalman filter can provide accurate UQ for regularly observed nonlinear
dissipative systems, similar to what is considered here.

We study the method in the context of RD systems as they are canonical examples
of nonlinear, time-dependent phenomena in 2D, providing the necessary increase
in state dimension from 1D examples considered previously. Systems often consist
of coupled sets of equations, further increasing the dimensionality. They are an
appropriate candidate for finite element methods, due to being
parabolic~\cite{thomeeGalerkinFiniteElement2006}, with potentially complex
geometries, and are an appropriate test-case for filtering; the dynamics are not
highly nonlinear, so the LR-ExKF can be applied without modification. However
the methodology is not restricted to RD systems and can be applied to any
nonlinear, time-dependent PDE for which finite element discretisations are
appropriate. The method can also be applied to linear PDEs, where the algorithm
gives the standard Kalman filter~\cite{kalmanNewApproachLinear1960}, and to
static PDEs, in which the algorithm reduces to the iterative application of
Bayes theorem.

The contribution of this paper is a scalable statFEM that can be applied to
systems with a high state dimension. This is presented, with a self-contained
introduction to statFEM, in Section~\ref{sec:statfem}, in the context of RD
equations. We demonstrate the method in Section~\ref{sec:case-studies}, on a
series of examples. The first is a 1D example with experimental data, and we
show that the LR-ExKF is able to accurately reproduce the full-rank ExKF, with
relative errors of $\order(10^{-6})$ on the mean and $\order(10^{-5})$ on the
variance. The next two examples deal with synthetic data, using the
Oregonator~\cite{fieldOscillationsChemicalSystems1972,tysonTargetPatternsRealistic1980},
a 2D system of two coupled PDEs, and we demonstrate scalability, taking
the state dimension up to $132,098$ degrees-of-freedom (DOFs). Examples show that the
LR-ExKF accurately reproduces the data generating process under misspecified
initial conditions, with relative errors of order $\mathcal{O}(10^{-2})$ in the
posterior mean. We also use the effective
rank~\cite{gottwaldMechanismCatastrophicFilter2013,patilLocalLowDimensionality2001}
to verify the effective dimensionality of the covariance matrix, ensuring that
the number of modes chosen is adequate. Section~\ref{sec:conclusion} concludes
the paper, and the Appendix includes two additional examples of interest to
practitioners, dealing with parameter estimation
(\ref{sec:app-parameter-estimation}) and filter divergence
(\ref{sec:app-divergence}). We also provide code to reproduce all results and
figures in this paper, available at
\url{https://github.com/connor-duffin/low-rank-statfem}.

\section{Methodology: low-rank statFEM}
\label{sec:statfem}

We introduce the methodology using a general RD system, given by
the semilinear parabolic PDE
\begin{equation}
  \begin{gathered}
    \partial_t u = \kappa \nabla^2 u + r(u) + \xi, \quad \text{ in } \Omega, \\
    \nabla u \cdot \mb{n} = 0, \quad \text{ on } \partial \Omega, \\
    u(\mb{x}, 0) = u_0, \quad u := u(\mb{x}, t) \in \mathbb{R}, \\
    \mb{x} \in \Omega \subset \mathbb{R}^d, \quad t \in [0, T].
  \end{gathered}\label{eq:rd-determ}
\end{equation}
Equation~\ref{eq:rd-determ} describes the evolution of a system state (e.g.
concentration of chemical species) which diffuses throughout the medium and has
nonlinear interactions. All parameters of Equation~\eqref{eq:rd-determ} above
are assumed known and are denoted by $\Lambda$. In this section, without
loss of generality, we deal with single-state equations with $u \in \mathbb{R}$.
Extensions to systems of equations with states $\bm{u} \in \mathbb{R}^s$ are
covered in Section~\ref{sec:case-studies}.

A $\GP$, $\xi := \xi(\mb{x}, t)$, is introduced inside of the PDE to represent
prior uncertainty in the model specification:
\begin{equation}
  \label{eq:gp-forcing}
  \xi \sim \GP(0, \delta(t - t')\cdot k_\theta(\mb{x}, \mb{x}')).
\end{equation}
The spatial covariance kernel $k_\theta(\cdot, \cdot)$ encodes
information on model error; we use the squared exponential
kernel~\cite{williamsGaussianProcessesMachine2006}
\[
  k_\theta(\mb{x}, \mb{x}') = \rho^2 \exp
  \left( -\frac{\lVert \mb{x} - \mb{x}' \rVert_2^2}{2 \ell^2}  \right)
\]
under the \textit{a priori} assumption that model errors are spatially smooth.
Hyperparameters $\bm{\theta} = (\rho, \ell)$ respectively parameterise the
variance and correlation length-scales of the model error, and are either set to
fixed values \textit{a priori} or estimated using maximum-a-posteriori
methods~\cite{murphyMachineLearningProbabilistic2012}. Hyperparameter estimation
is discussed after the algorithm presentation and is illustrated
in~\ref{sec:app-parameter-estimation}.

Delta correlations in time have the implication that
$\xi$ is the weak derivative of a function-valued Wiener process such
that $\xi(\mb{x}, t + \Dt) - \xi(\mb{x}, t) \sim \GP(0, \Dt k_\theta(\mb{x}, \mb{x}'))$.
We assume that $\xi(\mb{x}, 0) \equiv 0$, and thus
$\xi(\mb{x}, 1) \sim \GP(0, k_\theta(\mb{x}, \mb{x}'))$.
Spatially, the process inherits the regularity as implied by
$k_\theta(\cdot, \cdot)$, which in this work is assumed to give
$\xi(\mb{x}, \cdot) \in L^2(\Omega)$. For further details we refer
to~\cite{dapratoStochasticEquationsInfinite2014}.

Spatial discretisation proceeds through multiplying by testing functions
$\varphi \in H^1(\Omega)$ and integrating over the problem domain. This gives the
semidiscrete weak form
\[
  \langle \partial_t u, \varphi \rangle
  + \kappa \mathcal{A}(u, \varphi)
  = \langle r(u), \varphi \rangle
  + \langle \xi, \varphi \rangle, \quad \forall \varphi \in H^1(\Omega),
\]
where
$\mathcal{A}(u, \varphi) = \int_\Omega \nabla u \cdot \nabla \varphi \, \dee \mb{x}$,
$\langle f, g \rangle = \int_\Omega f \, g \, \dee \mb{x}$: these are
respectively the bilinear form induced by the Laplacian operator and the
$L^2(\Omega)$ inner product. Neumann boundary conditions result in the boundary
terms dropping out of the weak form.

The finite element mesh is given by subdividing the domain $\Omega$ into the
triangulation $\Omega_h \subseteq \Omega$ with vertices $\{\mb{x}_j\}_{j =
  1}^{n_u}$, where the maximal length of the sides of the triangulation is given
by $h$. The polynomial basis functions $\{\phi_j\}_{j = 1}^{n_u}$ are
defined on the mesh, having the property that $\phi_j(\mb{x}_i) = \delta_{ij}$.
Letting $V_h  = \mathrm{span}\{\phi_i\}_{i = 1}^{n_u}$, so that
$V_h \subset H^1(\Omega)$, we write the approximation
$u_h(\mb{x}, t) = \sum_{i = 1}^{n_u} u_{i}(t) \phi_i(\mb{x})$. In the
finite dimensional space $V_h$ the weak form is
\[
  \langle \partial_t u_h, \phi_j \rangle
  + \kappa \mathcal{A}(u_h, \phi_j)
  = \langle r(u_h), \phi_j \rangle
  + \langle \xi, \phi_j \rangle,
  \quad j = 1, \ldots, n_u,
\]
which corresponds to the $n_u$-dimensional SDE for the FEM coefficients
\[
  \mb{M} \, \dee \mb{u} + \kappa \mb{A} \mb{u} \, \dee t
  = \tilde{\mb{r}}(\mb{u}) \, \dee t + \dee \bm{\beta},
\]
where $\mb{M}_{ji} = \langle \phi_i, \phi_j \rangle$,
$\mb{A}_{ji} = \mathcal{A}(\phi_i, \phi_j)$,
$\tilde{\mb{r}}(\mb{u})_j = \langle r(u_h), \phi_j \rangle $,
and
$\mb{u}(t) = (u_1(t), \ldots, u_{n_u}(t))^\top$.
The additive noise process $\bm{\beta}(t) := (\beta_1(t), \ldots \beta_{n_u}(t))^\top$
has the increments
\[
  \bm{\beta}(t + \Dt) - \bm{\beta}(t) \sim \mathcal{N}(\mb{0}, \Dt \mb{G}_\theta),
  \quad
  \mb{G}_{\theta,{ij}} = \langle \phi_i,
  \langle k_\theta(\cdot, \cdot), \phi_j \rangle \rangle.
\]

In the absence of additive noise (i.e., for the ODE case), under bounded
derivative conditions on $r$, solutions will be unique, by the Picard-Lindel\"of
theorem~\cite{thomeeGalerkinFiniteElement2006}. We use Crank-Nicolson for time
integration; writing
$\mb{u}_{n} = (u_1(n \Dt), \ldots, u_{n_u}(n \Dt))^\top$,
for timestep size $\Dt > 0$, yields a fully discrete system accurate
to order $\mathcal{O}(h^2 + \Dt^2)$
\begin{equation}
  \mb{M} \left( \mb{u}_{n} - \mb{u}_{n - 1} \right)
  + \Dt \kappa \mb{A} \mb{u}_{n - 1/2}
  = \Dt \tilde{\mb{r}}(\mb{u}_{n - 1/2})
  + \mb{e}_{n - 1}, \quad \mb{e}_{n - 1} \sim \mathcal{N}(\mb{0}, \Dt \mb{G}_\theta),
  \label{eq:statfem-rd-prior-discrete}
\end{equation}
in which $\mb{u}_{n - 1/2} = (\mb{u}_{n} + \mb{u}_{n - 1}) / 2$. In practice any
appropriate time discretisation can be implemented and the choice will be
problem-dependent. In this paper Crank-Nicolson discretisations ensure
stability, motivated by the example given in~\ref{sec:app-divergence}. In this
example, it is seen that for sufficiently nonlinear dynamics the resultant
filtering algorithm may diverge when the time discretisation is of
implicit-explicit-type~\cite{ascherImplicitExplicitMethodsTimeDependent1995}.
The divergence is avoided when Crank-Nicolson is used.

Given a fixed set of hyperparameters $\bm{\theta}$, the Euler-Maruyama
discretisation~\cite{kloedenpeterNumericalSolutionStochastic1992} of
Equation~\eqref{eq:statfem-rd-prior-discrete} draws sample paths from the
measure $p(\mb{u}_n \given \bm{\theta}, \Lambda)$ for each $n = 1, \ldots, n_t$.
These samples are drawn from a physics-motivated prior distribution, which can
be updated, via Bayes theorem, to give a posterior distribution over the FEM
coefficients. Model misspecification can then be directly corrected for through
computing the posterior distribution.

The data generating process is assumed to be
$\mb{y}_n = \mb{H} \mb{u}_n + \bm{\eta}_n$. The data $\mb{y}_n \in \mathbb{R}^{n_y}$ are
assumed to be corrupted by an independent measurement noise process, $\bm{\eta}_n \sim
\mathcal{N}(0, \sigma^2 \mb{I}_{n_y})$, and are observed via the linear
observation operator $\mb{H} : \mathbb{R}^{n_u} \to \mathbb{R}^{n_y}$. For each
$n$, this defines the likelihood
$p(\mb{y}_n \given \mb{u}_n, \sigma) = \mathcal{N}(\mb{H} \mb{u}_n, \sigma^2 \mb{I}_{n_y})$.
The posterior distribution
$p(\mb{u}_n \given \mb{y}_{1:n}, \bm{\theta}, \sigma, \Lambda)$, with $\mb{y}_{1:n} =
(\mb{y}_1, \ldots, \mb{y}_n)$, can be computed through nonlinear filtering
algorithms. In previous work the ExKF and EnKF were
used~\cite{duffinStatisticalFiniteElements2021}, however as discussed in the
introduction these algorithms can perform poorly in high-dimensional settings. To
circumvent this computational bottleneck we use the LR-ExKF algorithm,
which computes the Gaussian approximation
$p(\mb{u}_n \given \mb{y}_{1:n}, \bm{\theta}, \sigma, \Lambda) \approx
\mathcal{N}(\mb{m}_n, \mb{C}_n)$ from a low-rank approximation of the state
covariance matrix $\mb{C}_n = \mb{L}_n \mb{L}_n^\top$. This is constructed
with the leading eigenvalues and eigenvectors of the prior covariance matrix
$\mb{G}_\theta$ and the previous timestep covariance $\mb{C}_{n - 1}$. For similar
approaches
see~\cite{verlaanTidalFlowForecasting1997,gillijnsReducedRankTransform2006,rozierReducedOrderKalmanFilter2007,lawEvaluatingDataAssimilation2012}.

Assume that the distribution of the previous state is given by
$p(\mb{u}_{n - 1} \given \mb{y}_{1:n - 1}, \bm{\theta}, \sigma, \Lambda)
= \mathcal{N}(\mb{m}_{n - 1}, \mb{C}_{n -1})$, with
$\mb{C}_{n-1} = \mb{L}_{n-1} \mb{L}_{n-1}^\top$,
$\mb{L}_{n-1} \in \mathbb{R}^{n_u \times k}$. Furthermore, assume that a
low-rank square root of
$\mb{G}_\theta = \mb{G}_\theta^{1/2} \mb{G}_\theta^{\top/2}$ is also available, where
$\mb{G}_\theta^{1/2} \in \mathbb{R}^{n_u \times k'}$. Also note that
$D_n \tilde{\mb{r}} := \partial \tilde{\mb{r}}(\mb{u}_{n - 1/2}) / \partial
\mb{u}_n$, the Jacobian matrix.

For all timesteps $n = 1, \ldots, n_t$, the LR-ExKF proceeds as:
\begin{enumerate}
\item (Prediction step) Solve
  \[
    \mb{M} \left( \hat{\mb{m}}_n - \mb{m}_{n - 1} \right)
    + \Dt \kappa \mb{A} \hat{\mb{m}}_{n - 1/2}
    = \Dt \tilde{\mb{r}}(\hat{\mb{m}}_{n - 1/2}),
  \]
  for the prediction mean $\hat{\mb{m}}_n$ and compute the prediction covariance
  square root:
  \begin{gather*}
    \tilde{\mb{L}}_n =
      \Big[ \left(\mb{M} + \Dt (\kappa \mb{A}  + D_n \tilde{\mb{r}})
      \right)^{-1}
      \left(\mb{M} + \Dt D_{n - 1} \tilde{\mb{r}} \right) \mb{L}_{n - 1}, \\
      \Dt \left(\mb{M} + \Dt \kappa \mb{A} + \Dt D_n \tilde{\mb{r}} \right)^{-1} \mb{G}_\theta^{1/2} \Big]
  \end{gather*}
  noting that $\tilde{\mb{L}}_n \tilde{\mb{L}}_n^\top = \hat{\mb{C}}_n$ (the
  prediction covariance), and that
  $\tilde{\mb{L}}_n \in \mathbb{R}^{n_u \times (k + k')}$. Each column of
  $\tilde{\mb{L}}_n$ can be formed in parallel; see the following discussion for
  further details.
\item (Truncation step) Take the eigendecomposition $\tilde{\mb{L}}_n^\top
  \tilde{\mb{L}}_n = \mb{V}_n \bm{\Sigma}_n \mb{V}_n^\top$, where
  $\bm{\Sigma}_n = \mathrm{diag}(\varsigma_1, \ldots, \varsigma_{k + k'})$.
  Approximate $\hat{\mb{L}}_n = \tilde{\mb{L}}_n \left[\mb{V}_n\right]_{:, 1:k}$
  for the highest magnitude $k$ modes, so the prediction covariance is
  $\hat{\mb{C}}_n = \hat{\mb{L}}_n \hat{\mb{L}}_n^\top$.
\item (Update step) Update the mean:
  \begin{equation*}
    \mb{m}_n = \hat{\mb{m}}_n
    + (\mb{H} \hat{\mb{C}}_n)^\top \left(\mb{H} \hat{\mb{C}}_n \mb{H}^\top + \sigma^2 \mb{I}_{n_y}\right)^{-1}
    (\mb{y}_n - \mb{H} \hat{\mb{m}}_n).
  \end{equation*}
  And the covariance:
  \begin{gather*}
    \mb{L}_n = \hat{\mb{L}}_n \mb{R}_n, \\
    \mb{R}_n \mb{R}_n^\top = \mb{I}_k - \hat{\mb{L}}_n^\top \mb{H}^\top (\mb{H} \hat{\mb{C}}_n \mb{H}^\top + \sigma^2
    \mb{I}_{n_y})^{-1} \mb{H} \hat{\mb{L}}_n,
  \end{gather*}
  using a Cholesky decomposition or otherwise.
\end{enumerate}
If $k = k' = n_u$ then the LR-ExKF recovers the full ExKF exactly. If the datum
$\mb{y}_n$ is missing then only the prediction and truncation steps are
completed, to produce the posterior $p(\mb{u}_n \given \mb{y}_{1:n}, \bm{\theta},
\sigma, \Lambda) \equiv p(\mb{u}_n \given \mb{y}_{1:n - 1}, \bm{\theta}, \sigma, \Lambda)$.

\subsection*{Discussion}

First we compare the ExKF and LR-ExKF in terms of memory and operation counts.
For a general reference to these matrix computations we refer
to~\cite{golubMatrixComputations2013}. The standard ExKF requires that $\mb{C}_n$ and
$\mb{G}_\theta$ are stored in memory, which is $\mathcal{O}(n_u^2)$ in space. For
large DOF problems this is infeasible, and provides the main motivation for the
low-rank approach. If one employs the standard ExKF, though, then the prediction
step for the covariance matrix requires the solution of the sparse $n_u \times
n_u$ matrix $\mb{M} + \Dt (\kappa \mb{A} + D_n \tilde{\mb{r}})$, $2 n_u$ times for each
timestep. For large $n_u$ this becomes prohibitively expensive. If using a
direct solver is feasible, then this can be slightly mitigated by computing the LU
factorisation of $\mb{M} + \Dt (\kappa \mb{A} + D_n \tilde{\mb{r}})$, and
reusing the factors when solving for each column of $\tilde{\mb{L}}_{n - 1}$,
though this still requires running forward- and back-substitution $2n_u$ times
per timestep.

Furthermore, the update step requires the solution of the system
$\mb{H} \hat{\mb{C}}_n \mb{H}^\top + \sigma^2 \mb{I}_{n_y}$, $n_u$ times, for each
timestep, which requires $\mathcal{O}(n_y^3/3)$ operations to take the Cholesky
decomposition and $\mathcal{O}(n_y^2 n_u)$ to solve.

In comparison, LR-ExKF requires solving $\mb{M} + \Dt (\kappa \mb{A} + D_n
\tilde{\mb{r}})$, $k + k'$ times for each timestep, and is $\mathcal{O}((k + k')
n_u)$ in space. The truncation step incurs a cost of $\mathcal{O}((k + k')^3)$
to compute the eigendecomposition. The update step has a cost of
$\order(n_y^3/3)$ operations to take the Cholesky factorisation of $\mb{H}
\hat{\mb{C}}_n \mb{H}^\top + \sigma^2 \mb{I}_{n_y}$, but requires only $k$
solves, to give the cost $\order(n_y^2 k)$. Finally, the cost of the Cholesky
factor $\mb{R}_n$ is $\order(k^3/3)$. However, note that $k \ll n_u$, and in our
experience the cost of these decompositions is dwarfed by solving the system in
the prediction step.

Note also that each column of the prediction covariance square root
$\tilde{\mb{L}}_n$ is able to be computed in parallel. As each column requires
solving $\mb{M} + \Dt (\kappa \mb{A} + D_n \tilde{\mb{r}})$ this can result in
lower runtimes, especially when combined with, for example, algebraic
multigrid solvers \cite{saadIterativeMethodsSparse2003}. This parallelisation is
likely to be necessary when scaling up the LR-ExKF to larger systems than
those considered here, such as high-dimensional 3D models. This
is also similar to the parallelisation potential of the EnKF prediction step,
which has enabled widespread adoption of this
algorithm~\cite{evensenDataAssimilationEnsemble2009}.

It is also assumed that the covariance spectrum of $\mb{G}_\theta$ is rapidly
decaying so that the majority of the variance can be explained by $k' \ll n_u$
dominant modes. In this work we use the squared-exponential covariance function,
which is known to have a rapid spectral
decay~\cite{banerjeeParallelInversionHuge2013,zimmermannConditionNumberAnomaly2015}.
For efficient methods to decompose $\GP$ covariance matrices into their leading
eigenvalues we refer to
\cite{saatciScalableInferenceStructured2011,dietrichFastExactSimulation1997,solinHilbertSpaceMethods2020,charlierKernelOperationsGPU2021}.
As the filter is initialised with the $l^2$-optimal low-rank square-root and the
truncation step preserves the dominant modes of variation, it is thought that
this scheme is able to provide accurate UQ; we plan to verify this in future
work.

The low-rank approximation of the covariance matrix will lead to
underestimation. If we write the prediction covariance as $\hat{\mb{C}}_n +
\hat{\mb{C}}_{n,\mathrm{err}} $ then the norm of the discarded component is
given by $\Vert \hat{\mb{C}}_{n,\mathrm{err}} \Vert = \varsigma_{k + 1}$. This
is noted in \cite{gillijnsReducedRankTransform2006}, where it is also commented
that this underestimation could lead to similar problems as encountered in
ensemble Kalman filtering, such as catastrophic filter divergence
\cite{gottwaldMechanismCatastrophicFilter2013}. We have observed this when using
unstable time-integration schemes, in the spiral wave regime; we refer to
the~\ref{sec:app-divergence} for full details. There is additional symmetry to
the EnKF: both algorithms propagate a low-rank approximation to the covariance
square root ($\mb{L}_n$ in the LR-ExKF; the ensemble in the EnKF), and make the
Gaussian assumption in the update step. It is thought that the LR-ExKF is
similar to an EnKF where the ensemble members are chosen to optimally represent
the variance (in the $l^2$ sense), and the propagation is done through the tangent
linear model.

To avoid inverting $\mb{H} \hat{\mb{C}}_n \mb{H}^\top + \sigma^2 \mb{I}_{n_y}$,
when $n_y \gg k$, the Woodbury
matrix identity~\cite{rozierReducedOrderKalmanFilter2007,mandelEfficientImplementationEnsemble2006}
can be used
\begin{align*}
  \left(\mb{H} \hat{\mb{C}}_n \mb{H}^\top + \sigma^2 \mb{I}_{n_y} \right)^{-1}
  &= \left( \mb{H} \hat{\mb{L}}_n (\mb{H} \hat{\mb{L}}_n)^\top + \sigma^2 \mb{I}_{n_y} \right)^{-1} \\
  &= \frac{1}{\sigma^2}\left(
    \mb{I}_{n_y} + \mb{H} \hat{\mb{L}}_n \left(\sigma^2\mb{I}_k + (\mb{H} \hat{\mb{L}}_n)^\top (\mb{H} \hat{\mb{L}}_n) \right)^{-1}
    (\mb{H} \hat{\mb{L}}_n)^\top
    \right).
\end{align*}
The dense $k \times k$ symmetric positive definite matrix
$\sigma^2\mb{I}_k + (\mb{H} \hat{\mb{L}}_n)^\top (\mb{H} \hat{\mb{L}}_n)$ can be
solved using the Cholesky decomposition~\cite{golubMatrixComputations2013}.

$\GP$ hyperparameters $\bm{\theta}$ and noise standard deviation $\sigma$ can be
estimated through maximum-a-posteriori (MAP)
estimation~\cite{murphyMachineLearningProbabilistic2012}. To estimate parameters
we assume the time evolving structure $\bm{\theta} := \bm{\theta}_n$,
$\sigma := \sigma_n$ for all $n$, and
that $\bm{\theta}_n$ and $\sigma_n$ are independent across time. Note that
$\tilde{\mb{L}}_n \equiv \tilde{\mb{L}}_n(\bm{\theta}_n)$, and writing
$\bm{\theta}_{1:n} = (\bm{\theta}_1, \ldots, \bm{\theta}_n)$ and
$\sigma_{1:n} = (\sigma_1, \ldots, \sigma_{n})$, then the marginal likelihood is
\[
  p(\mb{y}_n \given \mb{y}_{1:n - 1}, \bm{\theta}_{1:n}, \sigma_{1:n}, \Lambda)
  \sim \mathcal{N} \left(
    \mb{H} \hat{\mb{m}}_n, \,
    \mb{H} \tilde{\mb{L}}_n(\bm{\theta}_n) \big(\mb{H} \tilde{\mb{L}}_n(\bm{\theta}_n)\big)^\top + \sigma_n^2 \mb{I}_{n_y}
  \right).
\]
The optimisation problem for the log-marginal-posterior is thus
\begin{equation*}
  \hat{\bm{\theta}}_n, \hat{\sigma}_n
  = \argmax_{\left(\bm{\theta}_n, \sigma_n\right)} \left\{
    \log p(\mb{y}_n \given \mb{y}_{1:n - 1}, \bm{\theta}_{1:n}, \sigma_{1:n}, \Lambda)
    + \log p(\bm{\theta}_n) + \log p(\sigma_n)
  \right\}.
\end{equation*}
In~\ref{sec:app-parameter-estimation}, we use the weakly informative priors
$\rho_n \sim \mathcal{N}_+(1, 1^2)$ and $\sigma_n \sim \mathcal{N}_+(0, 1^2)$,
reflecting the \textit{a priori} uncertainty in the optimal choices of these
hyperparameters. For the optimisation routine,
L-BFGS-B~\cite{nocedalNumericalOptimization2006} with positivity constraints, as
implemented in \texttt{SciPy}~\cite{virtanenSciPyFundamentalAlgorithms2020}, has
worked well in our experience.

\section{Case studies}
\label{sec:case-studies}

The methodology is demonstrated on three RD examples. The first is a verification
of the method on a 1D example of cell RD, using the experimental
data of~\cite{simpsonPracticalParameterIdentifiability2020}, and shows a
coherent synthesis of data with a prior physical model. It also confirms that
the low-rank filter accurately reproduces the full-rank filter, with relative
errors of $\order(10^{-6})$ for the mean, and $\order(10^{-4})$ for the
variance.

The next two examples are in 2D, using the
Oregonator RD model~\cite{fieldOscillationsChemicalSystems1972,tysonTargetPatternsRealistic1980}, with misspecificied initial conditions. The first
2D example demonstrates that observations of a single component of the system
can correct for misspecification on the unobserved component. The second 2D
example studies the affect of increasing the mismatch variance $\rho$ on filter
performance and also shows scalability, using a state dimension of $n_u =
132,098$. Conditioning on data corrects for misspecification, and in both cases
statFEM recovers the underlying data generating process to relative errors of
$\order(10^{-2})$. In these examples, running the full-rank filter is
prohibitively expensive, so we also use the effective rank of
$\hat{\mb{L}}_n$ as a measure for filter performance.

The supplement contains two additional studies, which discuss the parameter
estimation methodology, and a case of catastrophic filter divergence,
respectively. Code to run all examples is available on a public GitHub
repository~\footnote{Available at
\protect\url{https://github.com/connor-duffin/low-rank-statfem}.}, with all finite element
discretisations done in
\texttt{Fenics}~\cite{loggAutomatedSolutionDifferential2012}.

\subsection{Experimental data: verification}
\label{sec:cell-example}

We consider a system of two coupled nonlinear RD equations,
which model the densities of two different cell populations, as cells react with
one another and diffuse throughout the
domain, as discussed in~\cite{simpsonPracticalParameterIdentifiability2020}. The model is a
coupled system of two nonlinear RD equations, with stochastic forcing. The
combined system state is given by $\bm{w} = (u, v) \in \mathbb{R}^2$:
\begin{equation}
  \begin{cases}
    u_t = D u_{xx} - k_u u + 2 k_v v (1 - u - v) + \xi_u, \\
    v_t = D v_{xx} + k_u u - k_v v (1 - u - v) + \xi_v, \\
    u_x(0, t) = 0, \quad  u_x(1300, t) = 0, \\
    u := u(x, t), \quad x \in [0, 1300] \, \mu m, \quad t \in [0, 60] \, h. \\
  \end{cases}\label{eq:cell-rd-equations}
\end{equation}
Coefficients are set to $D = 700 \mu m^2 / h$, $k_u = 0.025$, $k_v =
0.0725$, with initial conditions
\[
  \begin{cases}
    u(x, 0) = v(x, 0) = 0, & x \in [400, 900], \\
    u(x, 0) = v(x, 0) = 0.055, & \text{otherwise}.
  \end{cases}
\]
In contrast to~\cite{simpsonPracticalParameterIdentifiability2020},
who linearly interpolate the data to give their initial conditions,
we assume the fixed piecewise initial conditions as above. Instead of
interpolating the data at time $t = 0$, we condition on it.
Equation~\eqref{eq:cell-rd-equations} is discretised using the standard linear
polynomial ``hat'' basis functions, for each component, on a regular mesh with
$200$ cells on the interval $[0, 1300]$. Crank-Nicolson is used for the time
discretisation, with timestep size $\Dt = 0.1$.

Hyperparameters of $\xi_u$ and $\xi_v$ are set to the same values, which are
constant across all times; this is to avoid propagating poor estimates through
the simulation. The covariance structure of Equation~\eqref{eq:gp-forcing}, is
used, and cross-correlations are assumed to be zero:
$\mathbb{E}(\xi_u \xi_v) \equiv 0$. Hyperparameters are set to $\rho = 2 \times
10^{-3}$ and $\ell = 100$.

Laboratory data of this system is contained in the supplementary information
of~\cite{simpsonPracticalParameterIdentifiability2020}, freely available online.
These data consist of observations of the two species at $4$ times,
$t_{\text{obs}} = 0, 16, 32, 48$ hours. The concatenated state,
$\mb{w}_n = {(\mb{u}_n^\top, \mb{v}_n^\top)}^\top$, gives the data
generating process
$\mb{y}_n = \mb{H} \mb{w}_n + \bm{\eta}_n$ at the observation times, with noise
$\bm{\eta}_n \sim \mathcal{N}(0, \sigma^2 \mb{I}_{n_y})$, $\sigma = 0.01$.

\begin{figure}[t]
  \centering \includegraphics[width=0.8\textwidth]{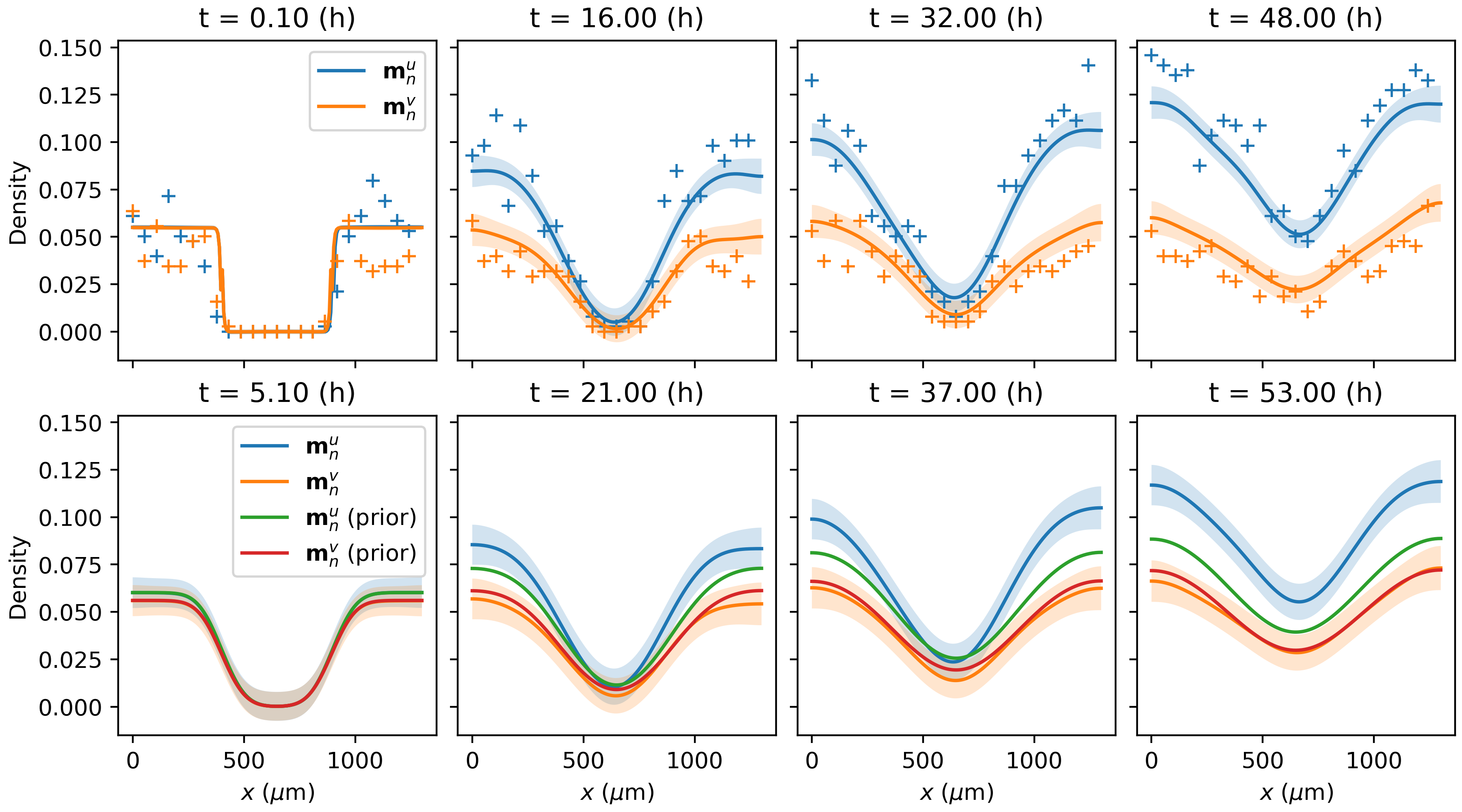}
  \caption{Observed data, posterior means, and 95\% posterior credible intervals
    for the four observed data times (top), and posterior means and 95\%
    posterior credible intervals for five hours after the observation times
    (bottom).}
  \label{fig:cell-post-means-vars}
\end{figure}

\begin{figure}[t]
  \centering \includegraphics[width=0.7\textwidth]{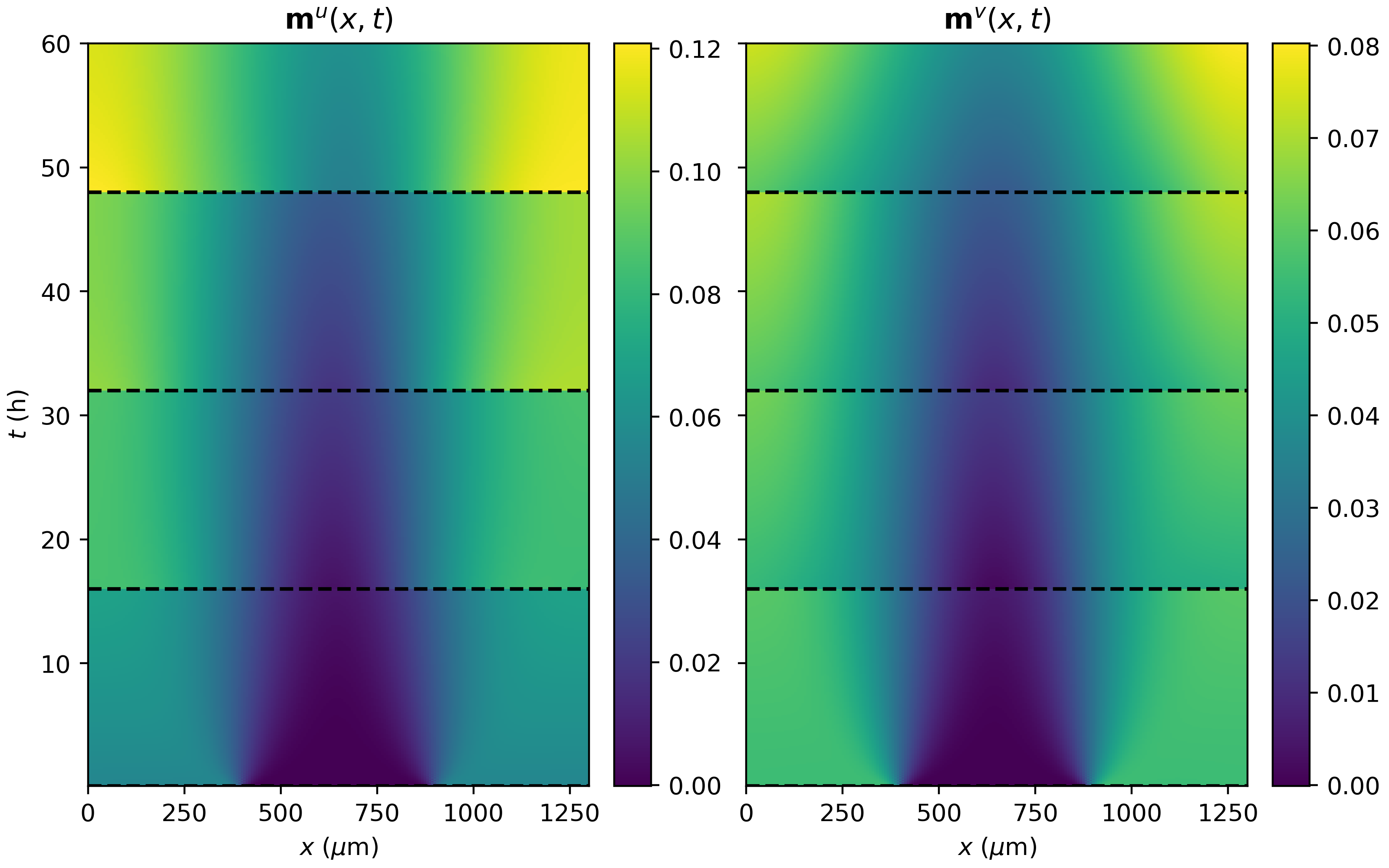}
  \caption{Space-time plot of the posterior means for each component. Times at
    which data are observed are shown with a dashed black line. On the $u$
    component there is an immediate correction from data conditioning, not
    present in the $v$ component: the prior model for $v$ is more accurate.}
  \label{fig:cell-surfaces}
\end{figure}

The posterior $p(\mb{w}_n \given \mb{y}_{1:n}, \bm{\theta}, \sigma, \Lambda) =
\mathcal{N}(\mb{m}_n, \mb{L}_n \mb{L}_n^\top)$ is computed with the LR-ExKF,
using $k = k' = 32$ modes for both the state covariance, and for the covariances
of the $\GP$s, $\xi_u$ and $\xi_v$. More than $99\%$ of the variance is retained in
the variance truncation at each timestep. The resulting posterior means $\mb{m}_n^u,
\mb{m}_n^v$, and $95\%$ posterior credible intervals are shown for both the
observation times and five hours after the observation times, in
Figure~\ref{fig:cell-post-means-vars}. For the $v$ component, there is little
discrepancy between the data and the prior assumed model, however for the $u$
component there is some degree of model mismatch, which conditioning on data can
partially account for. The posterior means for each component across the entire
space-time grid are also plotted in Figure~\ref{fig:cell-surfaces}, and
demonstrate the immediate effects of conditioning on data at the times at which
these data are observed.

The low- and full-rank ExKF's are now compared through the posterior means
and variances, in terms of the relative error
\[
  \frac{\lVert \mb{m}_n^u - \mb{m}_{n, LR}^u \rVert}{\lVert \mb{m}_n^u \rVert}, \quad
  \frac{\lVert \mathrm{var}(\mb{u}_n) - \mathrm{var}(\mb{u}_{n, LR}) \rVert}{\lVert
    \mathrm{var}(\mb{u}_n) \rVert},
\]
with the norm $\lVert \cdot \rVert$ taken as the standard Euclidean norm (the
$l^2$-norm).

\begin{figure}[t]
  \centering
  \includegraphics[width=0.6\textwidth]{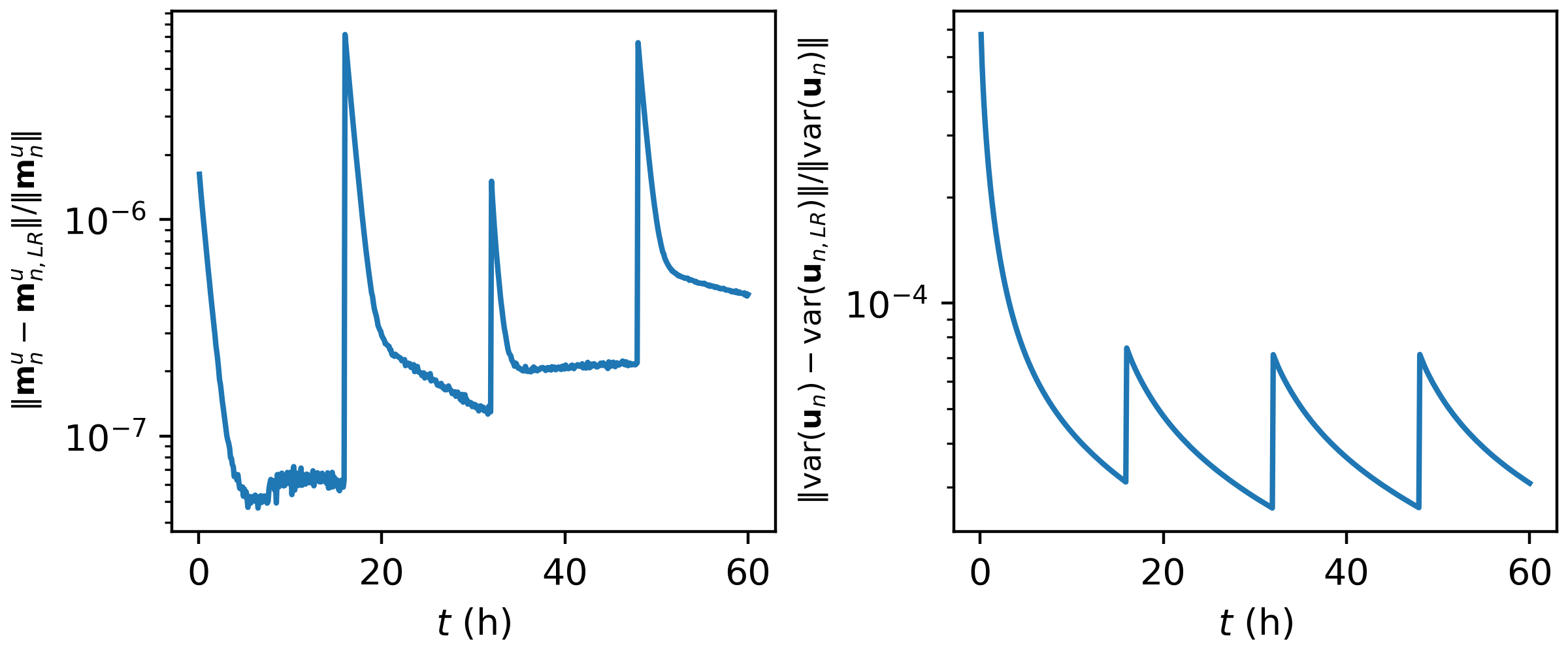}
  \caption{Relative error between the low-rank and full-rank ExKF, across all
    times, using $k = k' = 32$ modes, on the $u$ component.}
  \label{fig:cell-rel-error}
\end{figure}

\begin{figure}[t]
  \centering
  \begin{subfigure}[t]{0.48\textwidth}
    \centering \includegraphics[width=\textwidth]{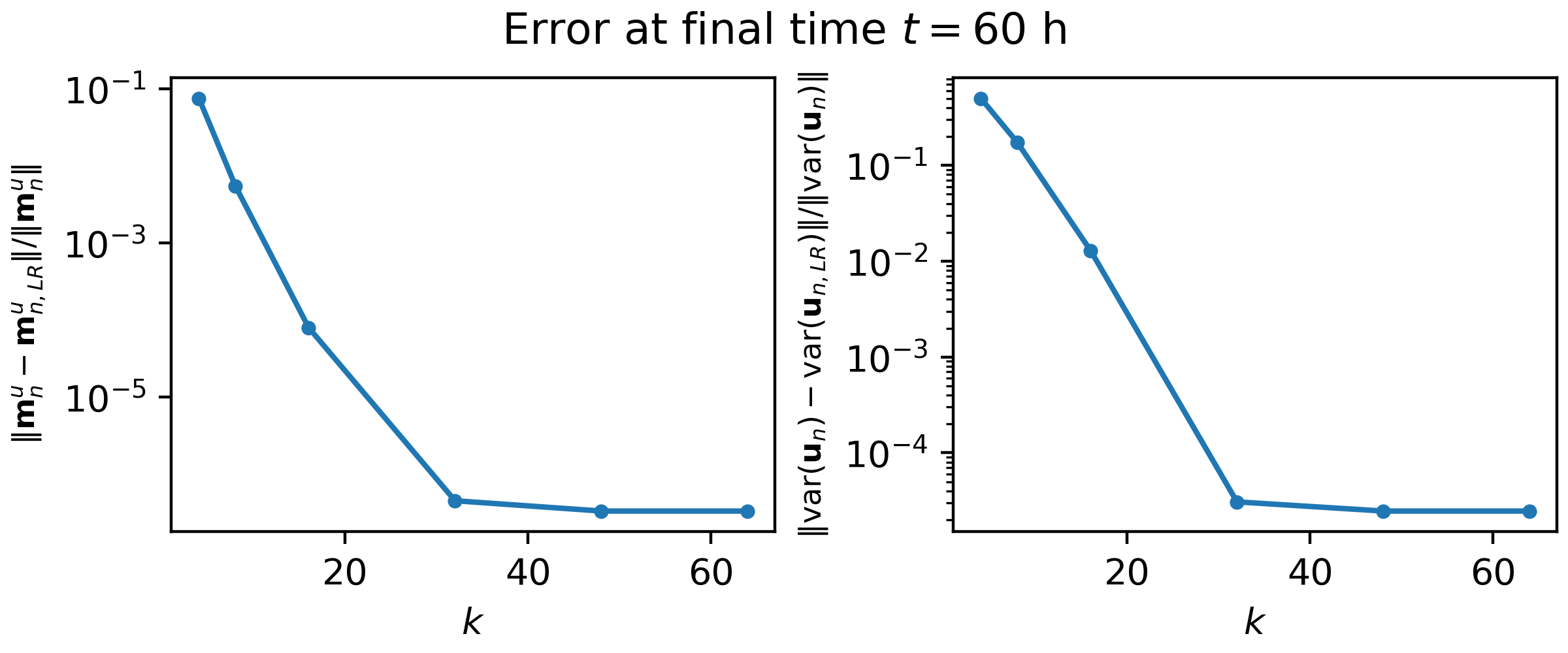}
    \caption{Relative error between the low-rank and full-rank ExKF, at the
      final time $t = 60$ h, as $k$ is increased ($k' = 32$ for all $k$).}
    \label{fig:cell-lr-modes-errors}
  \end{subfigure}
  ~
  \begin{subfigure}[t]{0.48\textwidth}
    \centering \includegraphics[width=\textwidth]{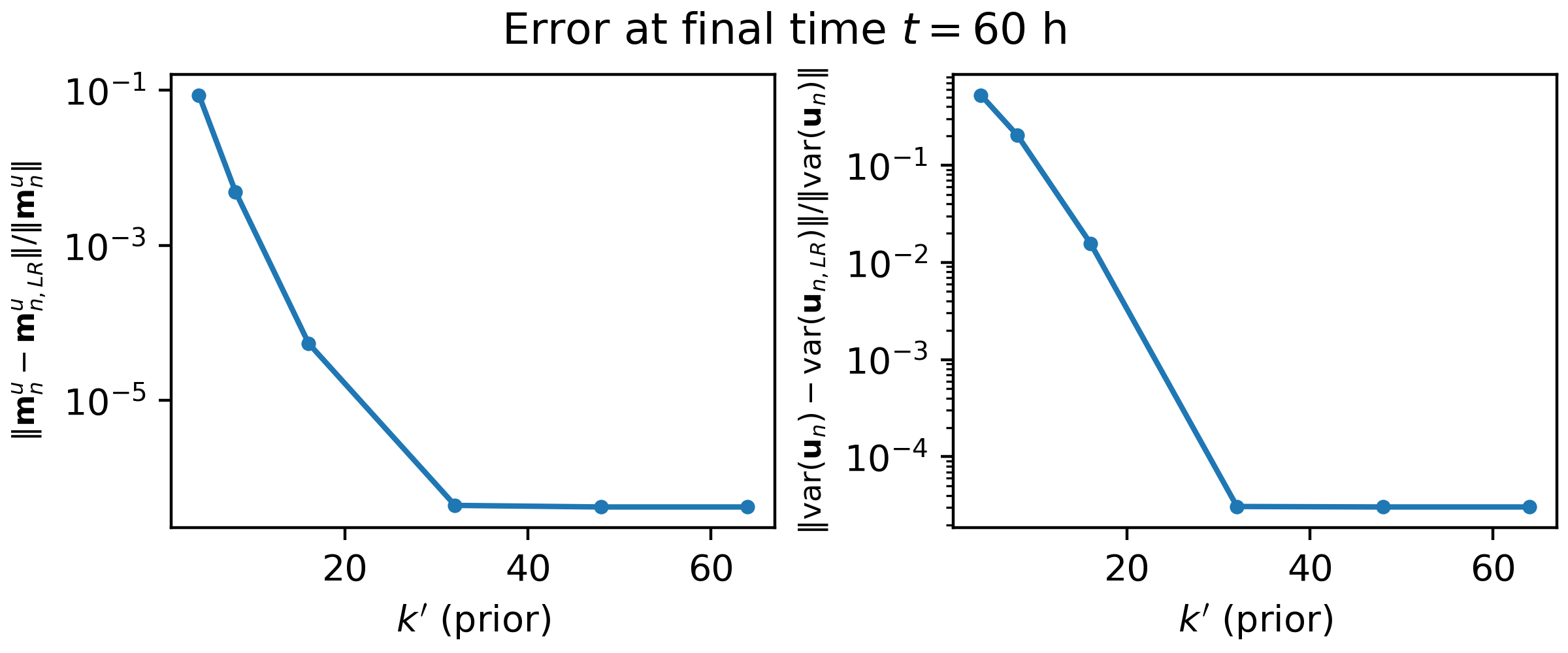}
    \caption{Relative error between the low-rank and full-rank ExKF, at the
      final time $t = 60$ h, as $k'$ is increased ($k = 32$ for all $k'$).}
    \label{fig:cell-lr-modes-errors-prior}
  \end{subfigure}
  \caption{Relative error comparisons between the low-rank and full-rank ExKF
    variants.}
  \label{fig:cell-errors-all}
\end{figure}

Shown in Figure~\ref{fig:cell-rel-error}, for a fixed number of
modes ($k = k' = 32$), the relative errors are small, at approximately $10^{-7}$
for the posterior mean, and $10^{-5}$ for the posterior variance. Sharp
increases are observed at the times at which data is observed, in both the
posterior mean and the posterior variance. The relative error at the end time of
the simulation is also shown in
Figure~\ref{fig:cell-errors-all}, as both $k$ and $k'$ are increased. In either
case, the errors are computed over $\{4, 8, 16, 32, 48, 64\}$ modes, with either
$k$ or $k'$ fixed to $32$. In each case, as the variable number of modes
increases the error decreases, with minor gains in the final error observed when
going from $32$ to $64$ modes.

As $k$ is increased (Figure~\ref{fig:cell-lr-modes-errors}), past $k = k' = 32$
there is a small increase in the accuracy of the filter, with a minor increase
from $k = 32$ to $k = 48$. This minor increase is thought to be due to the
inclusion of information available from the data, not present in the prior
alone. As $k'$ is increased (Figure~\ref{fig:cell-lr-modes-errors-prior}), past
$k' = k = 32$ there is visually no gain from adding in additional modes in the
prior, as these end up being truncated to $k = 32$. In both cases, increasing
from $48$ to $64$ modes shows little decrease in the posterior mean and variance
errors. These results suggest that there is a number of effective modes that
capture the dominant modes of variation; beyond some effective number of
dimensions, taking more modes does not yield significant gains in accuracy.

The low-rank approximation of the prior covariance matrix has a large affect on
the accuracy of the estimated posterior covariance matrix. Recall that in the
problem specification the uncertainty is induced via the additive $\GP$, $\xi$,
and no other sources of uncertainty are considered; if the covariance of $\xi$
is approximated to a sufficient degree then the low-rank approximation of the
filter is accurate.

This can be seen from the ExKF approximation of the prior measure, which
for exposition is considered in the case in which $\mb{G}_\theta$ is
rank-deficient but is stored fully in memory. This gives an iterative
approximation for the covariance matrix~\cite[Supplementary
Information]{duffinStatisticalFiniteElements2021}
\begin{equation*}
  \hat{\mb{C}}_n =
  \mb{J}_n^{-1}(\mb{J}_{n - 1} \mb{C}_{n - 1} \mb{J}_{n - 1}^\top + \mb{G}_\theta ) \mb{J}_n^{-\top},
\end{equation*}
where $\mb{J}_n$ is the Jacobian matrix of the FEM model with respect to
$\mb{u}_n$, evaluated at $(\hat{\mb{m}}_n, \mb{m}_{n - 1})$.
Given the recursive nature of computation and the fact that
$\mb{C}_0 = \mb{0}$ (due to the assumed initial conditions being exact) the
prior covariance is a sum of matrix products with $\mb{G}_\theta$ and the Jacobian
matrices, at each timestep. The prior covariance will be accurate if the
low-rank approximation of $\mb{G}_\theta$ is accurate. Empirically, this is also
seen for the posterior covariance. This phenomenon is analysed further in
\ref{sec:cell-prior-modes}, in which we investigate the errors on the mean and
the variance as both $k'$ and $\sigma$ are varied.

\subsection{Mismatch via initial conditions: spiral regime}

We now consider a two-dimensional example of misspecified initial conditions
with the Oregonator~\cite{fieldOscillationsChemicalSystems1972,tysonTargetPatternsRealistic1980},
a coupled PDE with state $\bm{w} = (u, v) \in \mathbb{R}^2$. Adding stochastic forcing
on the observed $v$-component gives the two-dimensional system
\[
  \begin{cases}
    u_t = \frac{1}{\varepsilon}\left( u(1 - u) - f v \frac{u - q}{u + q} \right)
    + D_u \nabla^2 u, \\
    v_t = u - v + D_v \nabla^2 v + \xi_v, \\
    \nabla u \cdot \mb{n} = 0, \quad \nabla v \cdot \mb{n} = 0, \quad x \in \partial \Omega, \\
    u := u(\mb{x}, t), \quad v := v(\mb{x}, t), \quad \mb{x} = (x_1, x_2) \in \Omega, \\
    \Omega = [0, 50] \times [0, 50], \quad t \in [0, 10].
  \end{cases}
\]
The Oregonator has been well-studied after being derived as a simplified model
for the chemical reaction kinetics of the Belousov-Zhabotinskii (BZ) reaction
\cite{fieldOscillationsChemicalSystems1972,tysonTargetPatternsRealistic1980,fieldOscillationsChemicalSystems1974,jahnkeChemicalVortexDynamics1989}.
It is a classical example of an activator-inhibitor system, sharing similar
behaviour to the Fitzhugh-Nagumo model in certain parameter
regimes~\cite{gongAntispiralWavesReactionDiffusion2003}. We study the Oregonator
in the excitable regime, setting $f = 2$, $q = 0.002$, and $\varepsilon = 0.02$.
Diffusion constants are set to $D_u = 1$, $D_v = 0.6$, The $\GP$ $\xi_v$ has the
covariance kernel of Equation~\eqref{eq:gp-forcing}, setting
$\bm{\theta} = (\rho, \ell) = (0.001, 5)$.

\begin{figure}[t]
  \centering
  \includegraphics[width=0.6\linewidth]{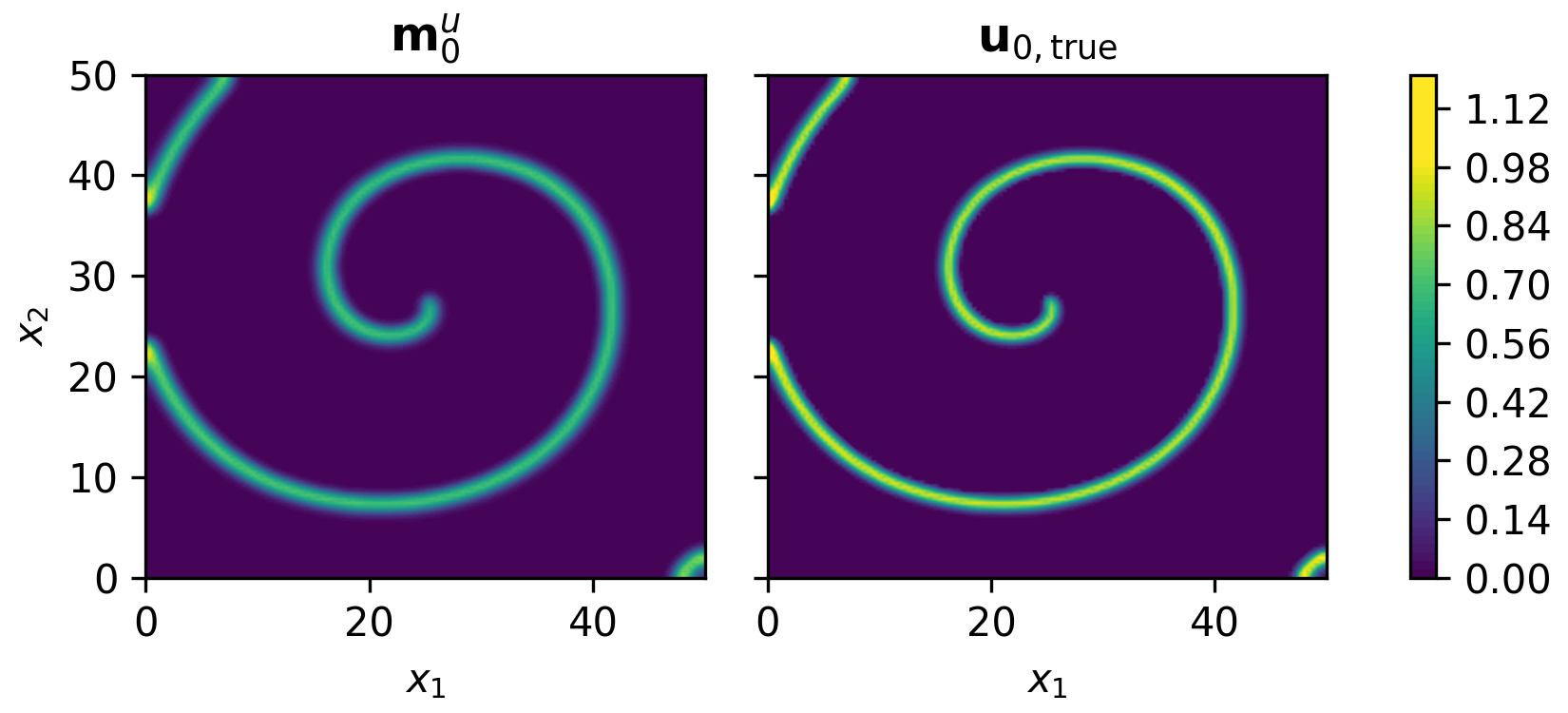}
  \caption{Initial conditions on the $u$ component. StatFEM posterior initial condition
    shown on the left, and the true initial condition shown on the
    right.}\label{fig:spiral-ic-compare}
\end{figure}

To set the initial conditions, we use the procedure
of~\cite{jahnkeChemicalVortexDynamics1989}. To induce some degree of mismatch,
initial conditions of the statFEM filtering posterior are set by pushing the
initial condition of the truth through a heat equation with the diffusion
coefficient $D_u = 1$ (i.e. the Oregonator sans reaction terms), for $t = 0.1$,
resulting in the amplitude of the initial conditions being dampened and
blurred (see Figure~\ref{fig:spiral-ic-compare}).

The spatial discretisation uses the standard linear polynomial hat functions,
over a regular mesh with $128 \times 128$ cells ($16,641$ nodes) so that the
total state dimension is $33,282$ DOFs. Due to the nonlinear terms
dominating the dynamics, Crank-Nicolson is used for time discretisation, with
timesteps $\Dt = 0.001$.

Data is observed on the slow $v$ component, at $1041$ locations inside the
discretised domain $\Omega_h$. The time between observations is $0.005$ (5
timesteps). Data is sparse in space, having approximately $3\%$ of the
state dimension observed at observation times, and the excitable $u$
is only updated through the observed $v$ values. As previous we concatenate the
discretised state vector into $\mb{w}_n = (\mb{u}_n^\top, \mb{v}_n^\top)^\top$ so that
the data generating process is $ \mb{y}_n = \mb{H} \mb{w}_n + \bm{\eta}_n$,
$\bm{\eta}_n \sim \mathcal{N}(0, \sigma^2 \mb{I}_{n_y})$. The measurement fidelity is
$\sigma = 0.01$.

We run the LR-ExKF to obtain the posterior
$p(\mb{w}_n \given \mb{y}_{1:n}, \bm{\theta}, \sigma, \Lambda) =
\mathcal{N}(\mb{m}_n, \mb{L}_n \mb{L}_n^\top)$
for all $n$, using $k = 250$ ($\approx 0.75\%$ of the state dimension) modes to
represent the state covariance $\mb{L}_n \mb{L}_n^\top$, and using $k' = 150$ modes to
approximate the covariance matrix of $\xi_v$, $\mb{G}_\theta$. When running the
filter, more than $99\%$ of the variance is retained at each truncation step.

\begin{figure}[t]
  \centering

  \begin{subfigure}{\textwidth}
    \centering
    \includegraphics[width=0.8\textwidth]{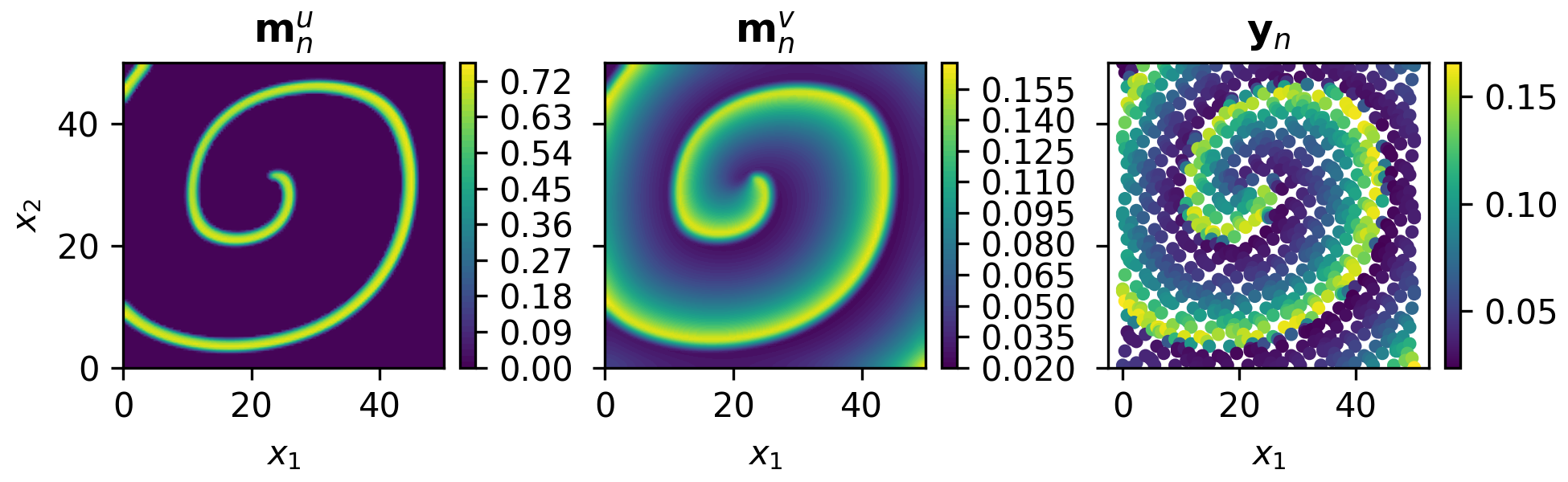}

    \includegraphics[width=0.6\textwidth]{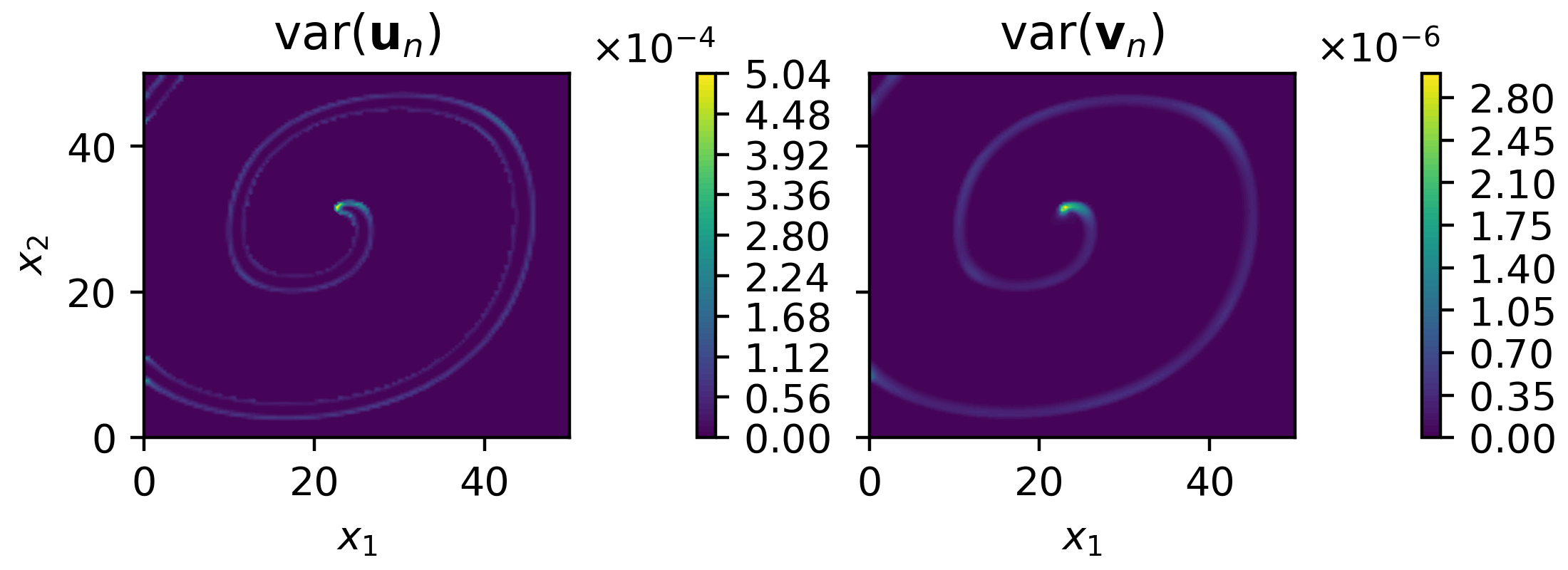}
    \caption{Top subplot: Posterior means $\mb{m}_u^n$ (top left), $\mb{m}_v^n$ (top
      centre), and observed data $\mb{y}_n$ (top right). Bottom subplot: posterior
      variances $\mathrm{var}(\mb{u}^n)$
      (bottom left) and $\mathrm{var}(\mb{v}^n)$ (bottom right) (diagonal of the covariance
      matrix $\mb{L}_n \mb{L}_n^\top$). All are reported for time $t = 5$.}
    \label{fig:spiral-post-means-vars}
  \end{subfigure}

  \begin{subfigure}{\textwidth}
    \centering
    \includegraphics[width=0.8\textwidth]{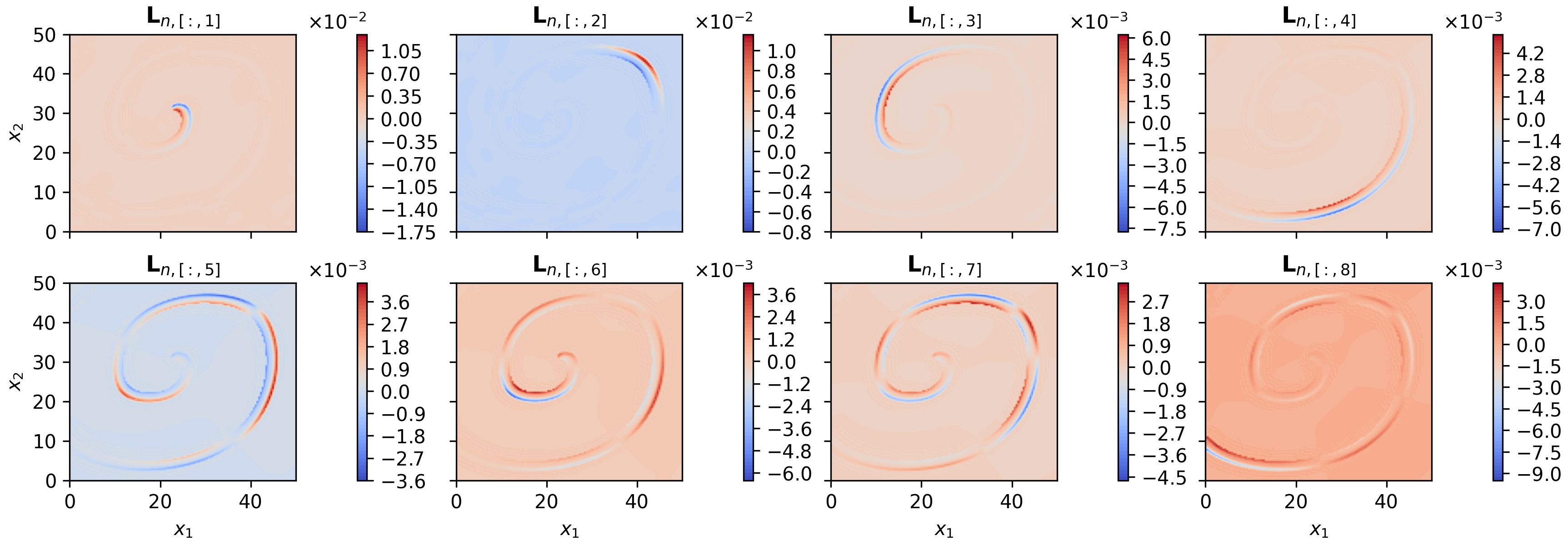}
    \caption{First eight columns of the posterior covariance square-root
      $\mb{L}_n$, for the $u$-component. These are the leading modes of
      variation at time $t = 5$.}
    \label{fig:spiral-post-modes}
  \end{subfigure}
  
  \caption{Posterior summary plots for time $t = 5$ for the Oregonator spiral
     example.}
  \label{fig:spiral-post-summary}
\end{figure}

To give a snapshot of results, the data, posterior mean, posterior variance, and leading
order modes are each shown in Figure~\ref{fig:spiral-post-summary}, at time $t =
5$. The dominant region of variance appears to be at the spiral tip, with
additional variation observed about the boundary of the spiral on the $u$
component. This hierarchy is seen in both the colour intensities of the variance
plots in Figure~\ref{fig:spiral-post-means-vars} and in the columns of $\mb{L}_n$, in
Figure~\ref{fig:spiral-post-modes}; for UQ, there is a hierarchy of variance
regions of decreasing importance.

\begin{figure}[t]
  \centering

  \begin{subfigure}[t]{0.6\textwidth}
    \includegraphics[width=\textwidth]{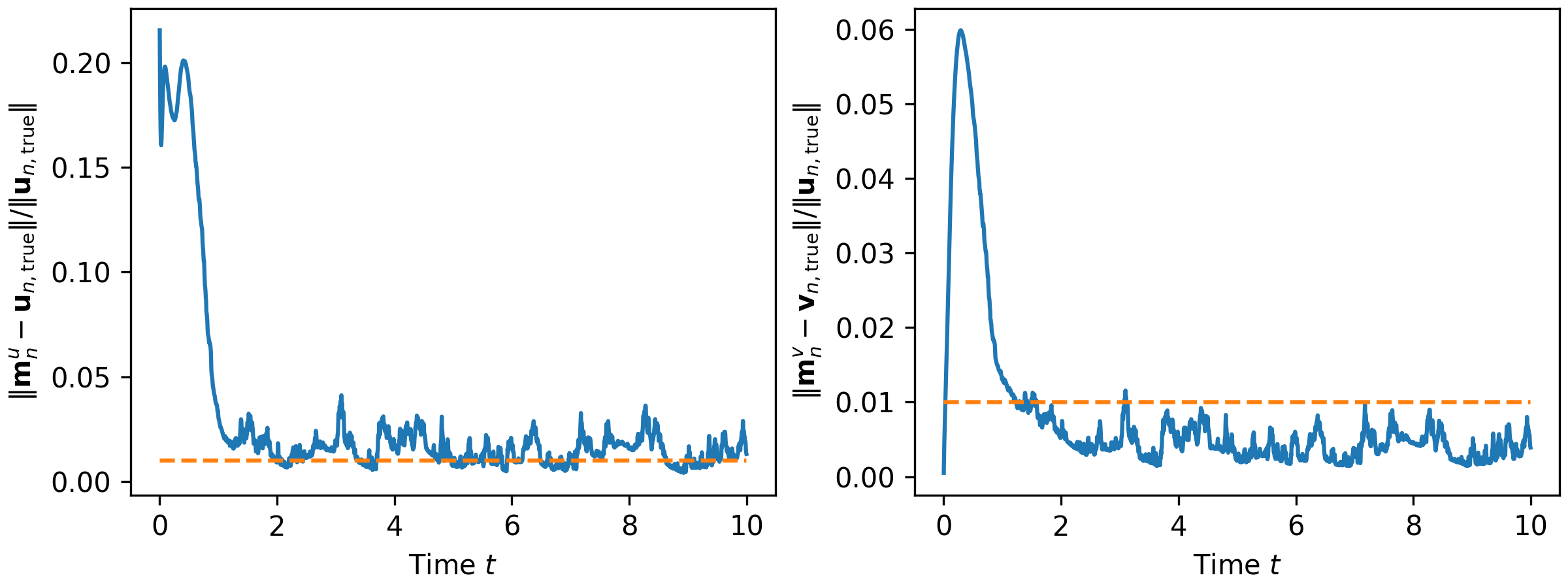}
    \caption{Relative errors
      ${\lVert \mb{m}_n^u - \mb{u}_{n, \text{true}} \rVert} / {\lVert \mb{u}_{n,
          \text{true}} \rVert}$, 
      and 
      ${\lVert \mb{m}_n^v - \mb{v}_{n, \text{true}} \rVert} / {\lVert \mb{v}_{n,
          \text{true}} \rVert}$, 
      for all times in the simulation.
    }
    \label{fig:spiral-rel-errors}
  \end{subfigure}
  \begin{subfigure}[t]{0.3\textwidth}
    \includegraphics[width=\textwidth]{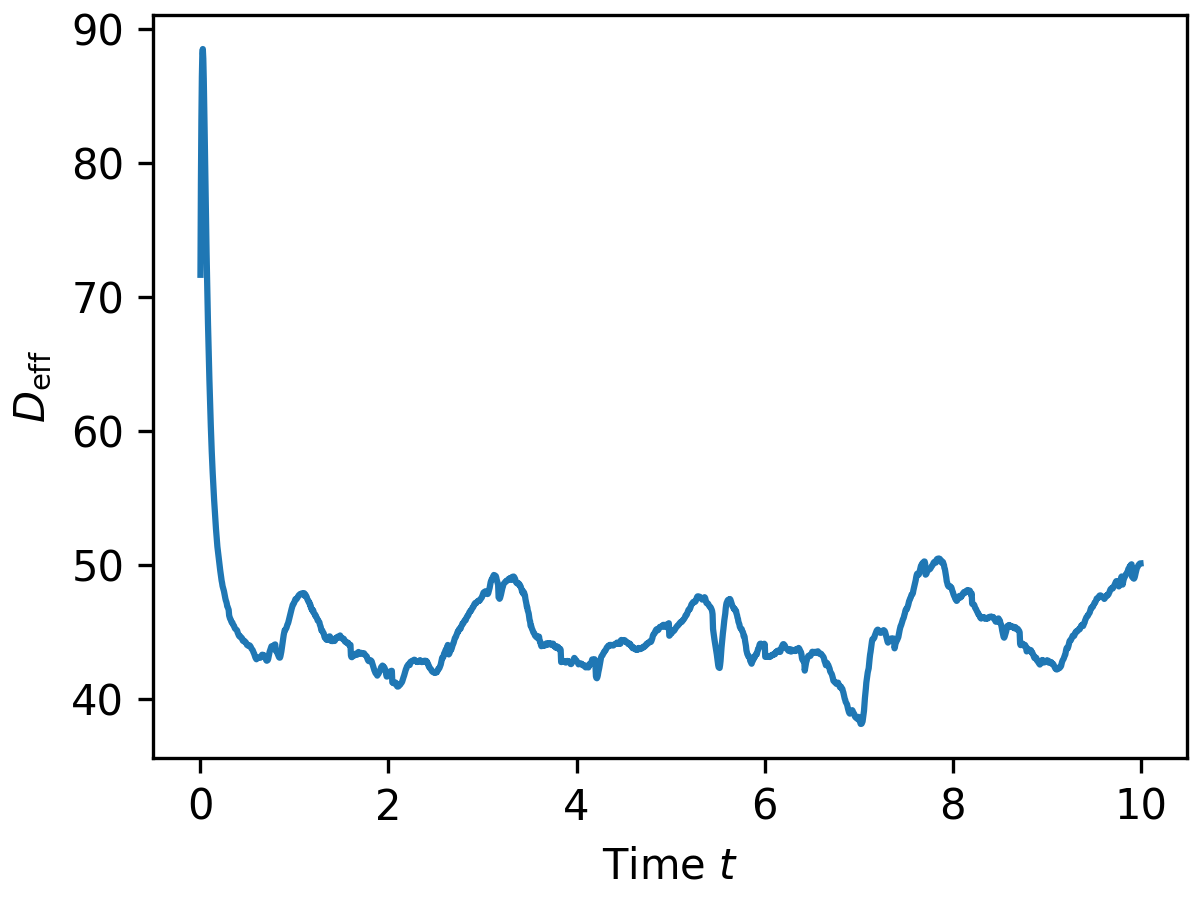}
    \caption{Effective rank of the prediction covariance square root $\hat{\mb{L}}_n$
      for all times in the simulation.}
    \label{fig:spiral-eff-rank}
  \end{subfigure}

  \caption{Diagnostic plots for the spiral wave example.}
  \label{fig:spiral-performance}
\end{figure}

Performance of the filter is verified through
computing the relative errors on the means $\mb{m}_n^u$, $\mb{m}_n^u$ against
the truth:
\[
  \frac{\lVert \mb{m}_n^u - \mb{u}_{n, \text{true}} \rVert}{\lVert \mb{u}_{n, \text{true}} \rVert}, 
  \quad
  \frac{\lVert \mb{m}_n^v - \mb{v}_{n, \text{true}} \rVert}{\lVert \mb{v}_{n, \text{true}} \rVert}.
\]
These are shown in Figure~\ref{fig:spiral-rel-errors}. After an initial period
of disparity, the statFEM mean $\mb{m}_n$ then closely tracks the truth, reaching a
stable configuration after this initial warm-up period. Observations can thus
correct for misspecification on the unobserved component, with small
($\order(10^{-2})$) relative errors in the mean of the statFEM posterior.

We also compute the effective rank
$D_\mathrm{eff}$~\cite{gottwaldMechanismCatastrophicFilter2013,patilLocalLowDimensionality2001}
of the prediction covariance matrix square-root $\hat{\mb{L}}_n$, using the
eigenvalues from the truncation step. We define the effective rank from the
eigenvalues (diagonal of $\bm{\Sigma}_n$)
$\varsigma_1 \geq \varsigma_2 \geq \cdots \geq \varsigma_k$ of the
$k \times k$ matrix $\hat{\mb{L}}_n^\top \hat{\mb{L}}_n$
\[
  D_\mathrm{eff} = \frac{\left(\sum_{i = 1}^k \sqrt{\varsigma_i}\right)^2}{\sum_{i = 1}^k \varsigma_i},
\]
which takes values $D_\text{eff} \in [1, \min\{k, n_u\}]$. This measures the
alignment of the columns of $\hat{\mb{L}}_n$ \cite{gottwaldMechanismCatastrophicFilter2013} and can be used to diagnose
problems (for example filter collapse), or verify performance (for example,
checking that $k$ and $k'$ are not over- or under-specified). For this example,
this is plotted in Figure~\ref{fig:spiral-eff-rank}, and appears to be stable
after an initial drop, further suggesting that the filter has reached a stable
configuration. The initial drop in the effective rank appears almost immediately
after the filter is started, whereas for the relative errors this is at time $t
= 2$. This suggests that reaching a stable configuration in the covariance is
perhaps necessary before the same occurs in the mean. The estimated $\Deff$ has
a mean value of approximately $\Deff \approx 45$, indicating that the choice of
$k = 250$ is perhaps excessive for this example.

\subsection{Mismatch via initial conditions: oscillatory regime}

\begin{figure}[t]
  \centering
  \includegraphics[width=0.6\linewidth]{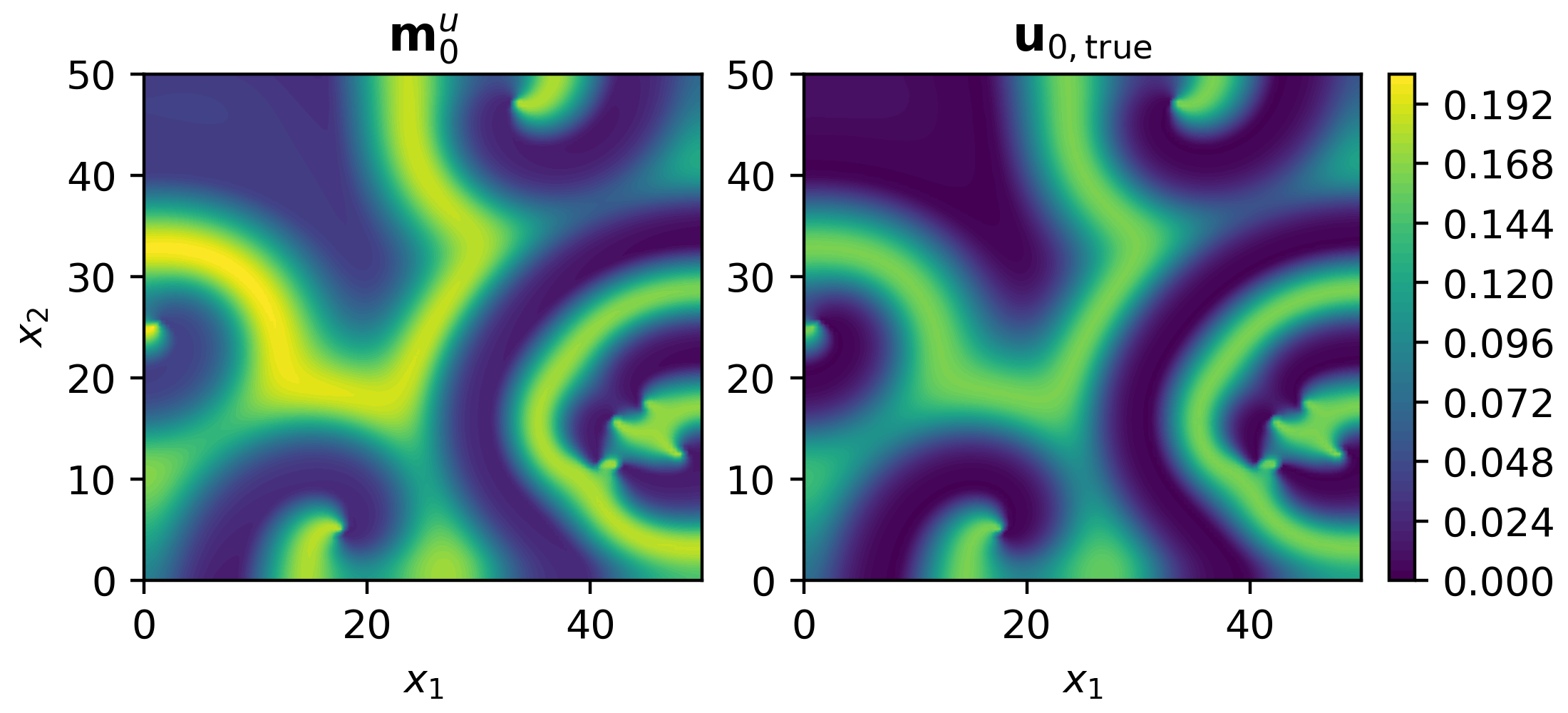}
  \caption{Initial conditions for $u$ component. StatFEM posterior initial
    condition shown on the left (same for each filter configuration) and the true
    initial conditions shown on the right.} \label{fig:ic-icu}
\end{figure}

In the oscillatory regime, the Oregonator has $f = 0.95$, $\varepsilon = 0.75$,
$q = 0.002$, and the diffusion coefficients
$D_u = D_v = 0.001$~\cite{gongAntispiralWavesReactionDiffusion2003}. Stochastic
forcing is now included on the $u$ component
\begin{align*}
  u_t &= \frac{1}{\varepsilon}\left( u(1 - u) - f v \frac{u - q}{u + q} \right)
        + D_u \nabla^2 u + \xi_u, \\
  v_t &= u - v + D_v \nabla^2 v,
\end{align*}
and the same spatio-temporal domain $\mb{x} \in \Omega = [0, 50] \times [0, 50]$,
$t \in [0, 10]$, is used. In this instance the initial conditions
$(u_{0, \mathrm{true}}, v_{0, \mathrm{true}})$ are perturbed via setting
\[
  u_0 = u_{0, \mathrm{true}}
  + \varkappa \left(1 + \sin\left( \frac{\pi x_1}{50} + \zeta_1  \right)
    \sin\left(\frac{\pi x_2}{50} + \zeta_2 \right) \right), \quad
  v_0 = v_{0, \mathrm{true}},
\]
where $\zeta_1, \zeta_2 \sim \mathcal{N}(0, 1^2)$, i.i.d. The amplitude $\varkappa$
is set to $0.02$ in our case studies, and for the true
$(u_{0, \mathrm{true}}, v_{0, \mathrm{true}})$, these are set from running a
pilot simulation for $100,000$ timesteps, where the pilot simulation has
randomly generated initial conditions $u, v \sim \mathrm{Unif}(0, 0.15)$. The
upper and lower bounds on the uniform distribution are determined from the
attractor of the corresponding Oregonator ODE. Both the perturbed and true
initial conditions are shown in Figure~\ref{fig:ic-icu}. This initial condition
is included in the accompanying
GitHub repository.

As previous the spatial discretisation uses linear polynomial basis functions,
on a refined mesh with $256 \times 256$ cells. Crank-Nicolson is used for the
time discretisation, with a timestep $\Dt = 10^{-2}$, observing data
$\mb{y}_n = \mb{H} \mb{w}_n + \bm{\eta}_n$ at every timestep up to
time $t = 10$. As previous, $\bm{\eta}_n \sim \mathcal{N}(0, \sigma^2 \mb{I}_{n_y})$,
with $\sigma = 10^{-2}$, and the $u$-component is observed at each timestep,
at $512$ observation locations. The posterior
$p(\mb{w}_n \given \mb{y}_{1:n}, \bm{\theta}, \sigma, \Lambda) =
\mathcal{N}(\mb{m}_n, \mb{L}_n \mb{L}_n^\top)$ is computed using the LR-ExKF
with $k = 128$ and $k' = 64$. The initial leading-order eigendecomposition of
the $\GP$ covariance matrix $\mb{G}_\theta$ is done using Lanczos
iterations~\cite{saadIterativeMethodsSparse2003} as implemented in
\texttt{Scipy}~\cite{virtanenSciPyFundamentalAlgorithms2020}, with the
matrix-vector products done on the GPU, implemented with
\texttt{KeOps}~\cite{charlierKernelOperationsGPU2021}.

We compare three filters, each with the $\GP$ covariance kernel of
Equation~\eqref{eq:gp-forcing}, which set $\rho \in \{10^{-2}, 10^{-3}, 2
\times 10^{-4} \}$, with $\ell = 10$ and $\sigma = 10^{-2}$ for each. In each
filter more than $99\%$ of the variance is retained at each truncation step. For
the remainder of this discussion these are referred to as the large-$\rho$ filter,
the moderate-$\rho$ filter, and the small-$\rho$ filter, respectively.

\begin{figure}[t]
  \centering

  \begin{subfigure}[t]{0.4\linewidth}
    \includegraphics[width=\linewidth]{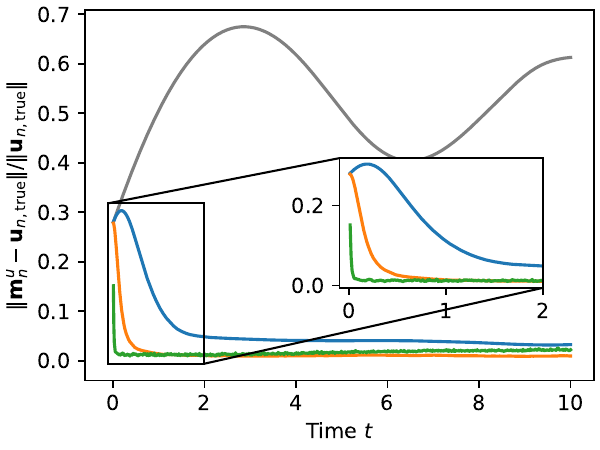}
    \caption{Relative $l^2$ errors for the large-$\rho$, moderate-$\rho$,
      small-$\rho$ filters, and the prior model.}
    \label{fig:ic-error}
  \end{subfigure}
  ~
  \begin{subfigure}[t]{0.4\linewidth}
    \includegraphics[width=\linewidth]{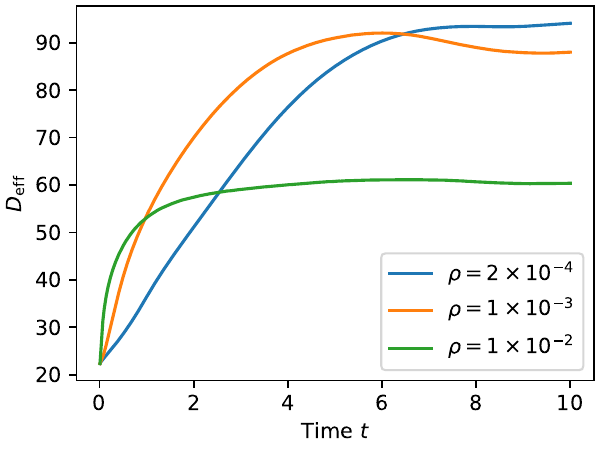}
    \caption{
      Effective rank of the prediction covariance square root $\hat{\mb{L}}_n$
      for the large-$\rho$, moderate-$\rho$, small-$\rho$ filters, and the prior
      model.
    }
    \label{fig:ic-eff-rank}
  \end{subfigure}
  
  \begin{subfigure}[t]{\linewidth}
    \centering
    \includegraphics[width=0.48\linewidth]{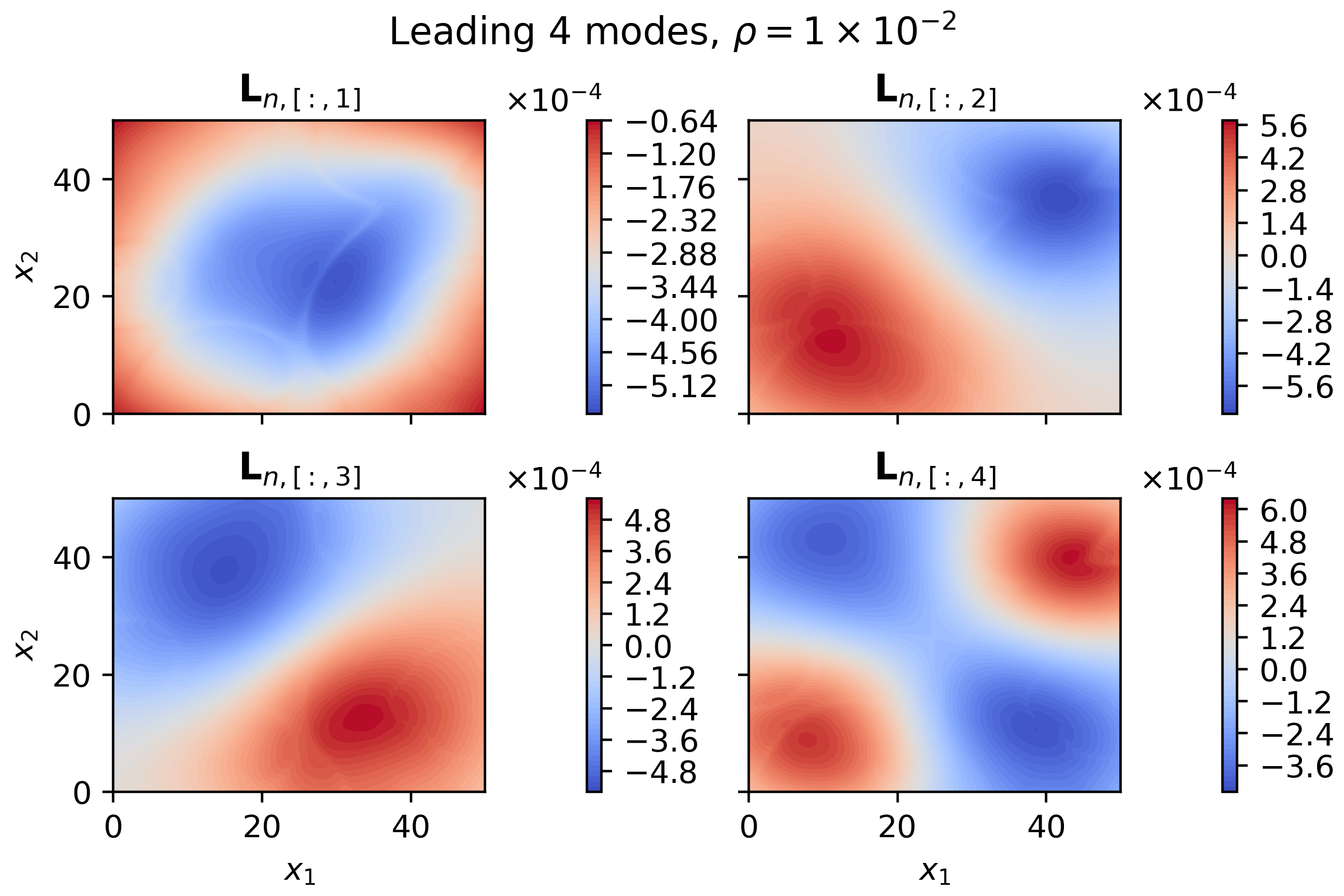}
    ~
    \includegraphics[width=0.48\linewidth]{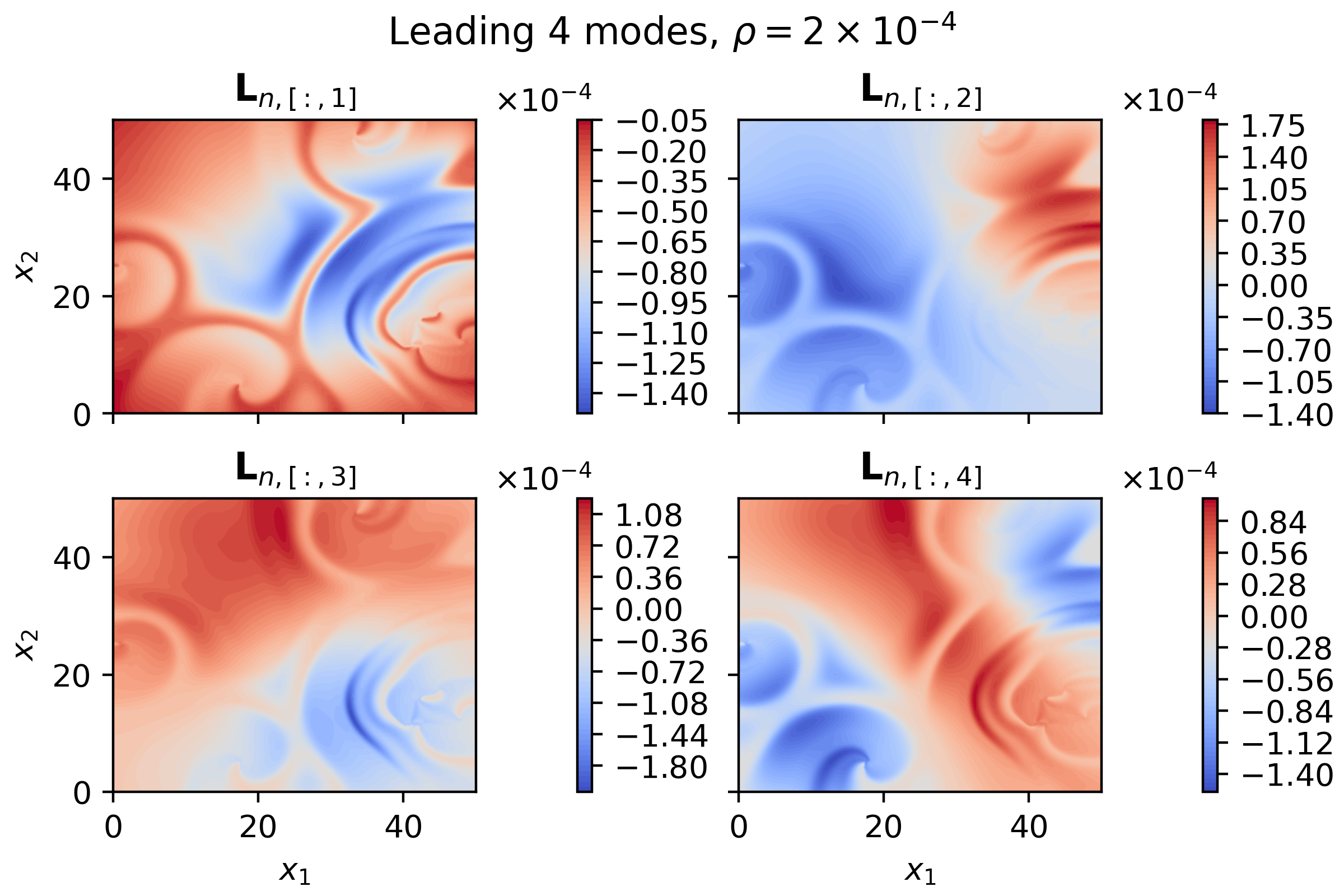}

    \caption{Leading four covariance modes (first four columns of the posterior
      covariance matrix square root $\mb{L}_n$) for the large-$\rho$ filter
      (left, \textcolor{tab_green}{green} (\textcolor{tab_green}{\rule[1.5pt]{0.4cm}{1pt}}) in
      Figures~\ref{fig:ic-error}~and~\ref{fig:ic-eff-rank}), and the small-$\rho$ filter
      (\textcolor{tab_blue}{blue} (\textcolor{tab_blue}{\rule[1.5pt]{0.4cm}{1pt}}) in
      Figures~\ref{fig:ic-error}~and~\ref{fig:ic-eff-rank}). Shown are the modes
      for the $u$-component.}
    \label{fig:ic-modes}
  \end{subfigure}

  \caption{Relative $l^2$ errors (top left), effective rank $\Deff$ (top right),
    and leading posterior covariance modes (bottom) at time $t = 10$, oscillatory
    mismatched initial condition. The small-$\rho$ is shown in
    \textcolor{tab_blue}{blue} (\textcolor{tab_blue}{\rule[1.5pt]{0.4cm}{1pt}}), the
    moderate-$\rho$ in \textcolor{tab_orange}{orange}
    (\textcolor{tab_orange}{\rule[1.5pt]{0.4cm}{1pt}}), the large-$\rho$ in
    \textcolor{tab_green}{green} (\textcolor{tab_green}{\rule[1.5pt]{0.4cm}{1pt}}), and the prior model
    in \textcolor{tab_gray}{grey} (\textcolor{tab_gray}{\rule[1.5pt]{0.4cm}{1pt}}).}
\label{fig:ic}
\end{figure}

Relative errors for the statFEM posterior means are shown alongside the statFEM
prior mean in Figure~\ref{fig:ic-error}. The small-$\rho$
filter shows the slowest misspecification correction compared to the others; the
filter is more certain of the model predictions and takes longer than the others to reach
a stable filtering configuration. It also results in the largest
relative error by the end of the simulation. The moderate $\rho$-filter,
however, despite slower initial misspecification correction, results in
the lowest relative error by $t = 10$. The large-$\rho$ filter displays the
most rapid convergence, yet results in relative errors increasing after the
initial correction. This is thought to be due to $\rho$ being equal to the
noise $\rho = \sigma = 10^{-2}$, which results in assimilation of spurious
noise perturbations. These results confirm the role of the $\rho$ hyperparameter
in specifying our UQ of the physical model. Higher variance implies less
certainty and more rapid corrections for model misspecification.

This trend of slower corrections, for more certain models, is also seen in the
effective rank $\Deff$ of the prediction covariance matrix $\hat{\mb{L}}_n
\hat{\mb{L}}_n^\top$, plotted for each of the filters in
Figure~\ref{fig:ic-eff-rank}. In terms of $\Deff$, the large-$\rho$ filter has
the most rapid convergence to a stable filtering regime, with the small-$\rho$
filter the slowest. Once in this stable filtering regime the small-$\rho$ filter
has a larger effective rank compared to the large-$\rho$ filter, and it is
posited that a higher effective rank implies a more complex covariance
structure. To check this the leading modes of the covariance matrix are plotted at
the end time $t = 10$, in Figure~\ref{fig:ic-modes}. The results suggests this
is the case, and we see that the modes for the small-$\rho$ filter display more
localised structures pertaining to the dynamics of the model, when compared with
those of the large-$\rho$ filter, which appear more similar to the
eigenfunctions of the $\GP$ covariance kernel $k_\theta(\cdot, \cdot)$.

This also merits another interpretation of the variance hyperparameter
$\rho$: that of controlling the weight of the \textit{a priori}
misspecification covariance matrix $\mb{G}_\theta$ in comparison to the (tangent
linear) dynamical evolution of the previous timestep covariance
$\mb{C}_{n - 1} = \mb{L}_{n - 1} \mb{L}_{n - 1}^\top$. In this case, due to the
complex spatial oscillations seen in the dynamical model, by decreasing the
weight of the simpler $\GP$ covariance, the more complex dynamical interactions
are present in the posterior covariance, resulting in an increased $\Deff$.

\section{Conclusions}
\label{sec:conclusion}

% what did we do: maths and methods
We present a low-rank extended Kalman filter algorithm for the statistical
finite element method that is highly scalable, by representing the
covariance matrix through its leading $k$ modes. Owing to the rapid spectral
decay of this posterior covariance matrix the LR-ExKF is able to provide a
sensible and interpretable UQ, retaining, in the given
examples, at least $99\%$ of the variance at each timestep. Through the use of
the Jacobian matrix (assembled from the Fr\'echet derivative of the weak form),
the statFEM LR-ExKF is straightforward to implement with standard finite element
libraries and has the potential for massive parallelisation in the bottleneck
prediction step.

% what did we do: experiments
We demonstrate, using experimental and synthetic data, that statFEM is able to
combine models and data in a coherent statistical framework, correcting for
model misspecification in various forms. Results in 1D, with experimental data,
show that the LR-ExKF can approximate the full-rank alternative, with small
relative errors in both the posterior mean and variance. These results also confirm
the efficacy of statFEM in the face of temporally sparse observation regimes.
Results with the 2D Oregonator system demonstrate scalability, with the statFEM
LR-ExKF correcting for model misspecification in the initial conditions, through
observing a single component of the underlying data-generating process. These
results are robust under different model parameter regimes, and, under
increasing the state dimension to $n_u = 132,098$. We also show that the
effective rank of the covariance matrix can verify the number of modes taken in
the low-rank approximation, providing an additional check of filter convergence.

This work provides a scalable statFEM that will enable the application of the
methodology to domains across the physical sciences. Our method is similar to
the EnKF, representing the covariance matrix through a low-rank approximation
(c.f.~\ref{sec:app-divergence}), though by using the leading eigenvalues and
eigenvectors we ensure that the UQ is empirically justified. An additional
source of uncertainty that is not considered in this work is that of the
hyperparameters, $\bm{\theta}$, which would ideally be marginalised
over~\cite{filipponePseudoMarginalBayesianInference2014}. The investigation of
this estimation approach would be of interest.

Additional extensions to this work include verifying that the LR-ExKF produces
accurate estimates of the true filtering posterior, and theoretical work to
check the accuracy and stability of the filter. Furthermore, when applying to
arbitrary time-dependent PDEs one may require the use of bespoke timestepping
schemes, going beyond the Crank-Nicolson scheme used in this paper. One may also
need to constrain the timesteps to ensure that the tangent linear approximation
to the covariance matrix does not induce too much error or to control for
possible filter divergence. An iterative extension may also be required
for sufficiently nonlinear PDEs~\cite{sakovIterativeEnKFStrongly2012b}.

\section*{Acknowledgements}

CD was supported by a Bruce and Betty Green Postgraduate Research Scholarship
and an Australian Government Research Training Program Scholarship at The
University of Western Australia. CD and EC were supported by the Australian
Research Council Industrial Transformation Research Hub Grant IH140100012, and
EC was supported by the Australian Research Council Industrial Transformation
Training Centre Grant IC190100031. CD and MG were supported by EPSRC grant
EP/T000414/1 and MG was supported by a Royal Academy of Engineering Research
Chair, and EPSRC grants EP/R018413/2, EP/P020720/2, EP/R034710/1, EP/R004889/1.

\bibliographystyle{elsarticle-num}
\bibliography{bibliography}

\newpage

\appendix

\section{Notes on numerics}
\label{sec:app-numerics}

All results in this paper were obtained on a workstation PC with an i7-6850k CPU,
equipped with 64GB of memory, and an nVidia GTX 1080 Ti GPU. All finite element
computations (i.e. matrix assembly and automatic differentiation for the
Jacobian) are handled with
\texttt{Fenics}~\cite{loggAutomatedSolutionDifferential2012}. At each timestep,
when solving the nonlinear system, we use the standard Newtons method, as
implemented in \texttt{PETSc}, using the LU decomposition to solve the linear
system at each Newton step. When performing the covariance propagation, i.e.
solving systems of the form
\[
  (\mb{M} + \Dt \kappa \mb{A} + \Dt D_n \tilde{\mb{r}})^{-1} \mb{X}, \quad
  \mb{X} \in \mathbb{R}^{n_u \times p},
\]
we again employ the LU factorization so as to reuse the factors.

We construct the matrix $\mb{G}_\theta$ through approximation, using
$\mb{G}_\theta = \mb{M} \mb{K}_\theta \mb{M}^\top$, for the mass matrix
$\mb{M}$ and the squared-exponential matrix
$\mb{K}_{\theta, ij} = k_\theta(x_i, x_j)$ (see the supplement of
\cite{duffinStatisticalFiniteElements2021} for further details). The low-rank
approximation $\mb{G}_\theta^{1/2}$ is computed using the leading $k'$ modes of
$\mb{K}_\theta$, which, depending on the problem at hand, are computed using vanilla
\texttt{Scipy} (on the CPU) or with \texttt{KeOps} (on the GPU). This gives
$\mb{G}_\theta^{1/2} = \mb{M} \mb{K}_\theta^{1/2}$.

\section{Influence of prior modes in the cell example}
\label{sec:cell-prior-modes}

As discussed in the main text (Section~\ref{sec:cell-example}), in the cell
example, the prior modes control the accuracy of the UQ given by the LR-ExKF.
Here we investigate this further, through checking the
errors in the posterior mean and variance as we vary the prior modes $k'$ and the
observational noise $\sigma$. For each $k'$ we fix the filter modes $k$ to $k =
k' + 16$, to allow for sufficient extra modes to capture information present in
the data, not included in the prior modes. Note that in this instance the
variation of $\sigma$ is only to investigate the computational performance of the
filtering method; the known noise value of $\sigma = 10^{-2}$ should be used (as it
is in the main text) when attempting to infer the system state from measurements.

\begin{figure}[!htb]
  \centering
  \includegraphics[width=0.8\linewidth]{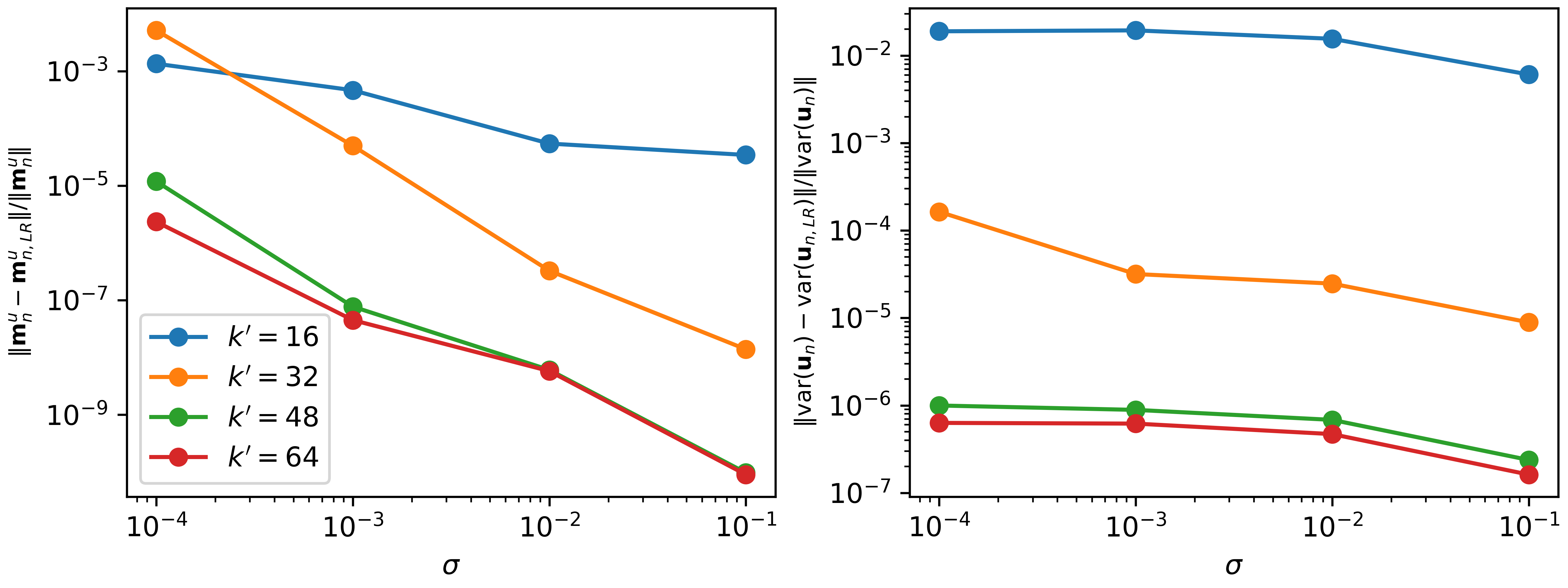}
  \caption{Relative errors on the posterior mean (left) and variance (right) as both the
    number of prior modes ($k'$) and observation noise ($\sigma$) is increased.}
  \label{fig:cell-lr-modes-sigma-errors}
\end{figure}

The results of running the filters for $k' \in \{16, 32, 48, 64\}$,
$\sigma \in \{10^{-4}, 10^{-3}, 10^{-2}, 10^{-1}\}$ are shown in
Figure~\ref{fig:cell-lr-modes-sigma-errors}. These suggest that as the noise
decreases (i.e., the data becomes more informative) then more modes are required
to accurately capture the variance. Despite the additional ``overhead modes'' on
$k = k' + 16$, gains are seen when increasing $k'$, confirming that the number
of prior modes have a large affect on the performance on the filter.
Errors in the posterior mean appear to increase more rapidly with the decreasing $\sigma$,
in comparison to the variance. It is posited that this discrepancy is seen due
to the posterior mean update using the full nonlinear dynamics when completing the
prediction step --- the covariance propagation, in comparison, uses the tangent
linear dynamics.

\section{Verification of parameter estimation}
\label{sec:app-parameter-estimation}

For this example, the Oregonator equations in the
oscillatory regime are used (recall $f = 0.95$, $\varepsilon = 0.75$, $q = 0.002$, and
the diffusion coefficients $D_u = D_v =
0.001$~\cite{gongAntispiralWavesReactionDiffusion2003}). Data is generated
according to a stochastic Oregonator with stochastic forcing on the $u$
component
\begin{align*}
  u_t &= \frac{1}{\varepsilon}\left( u(1 - u) - f v \frac{u - q}{u +
      q} \right) + D_u \nabla^2 u + \xi_u, \\
  v_t &= u - v + D_v \nabla^2 v,
\end{align*}
and we verify the proposed hyperparameter estimation routine of
Section~\ref{sec:statfem}.

\begin{figure}[t]
  \centering
  \includegraphics[width=0.4\linewidth]{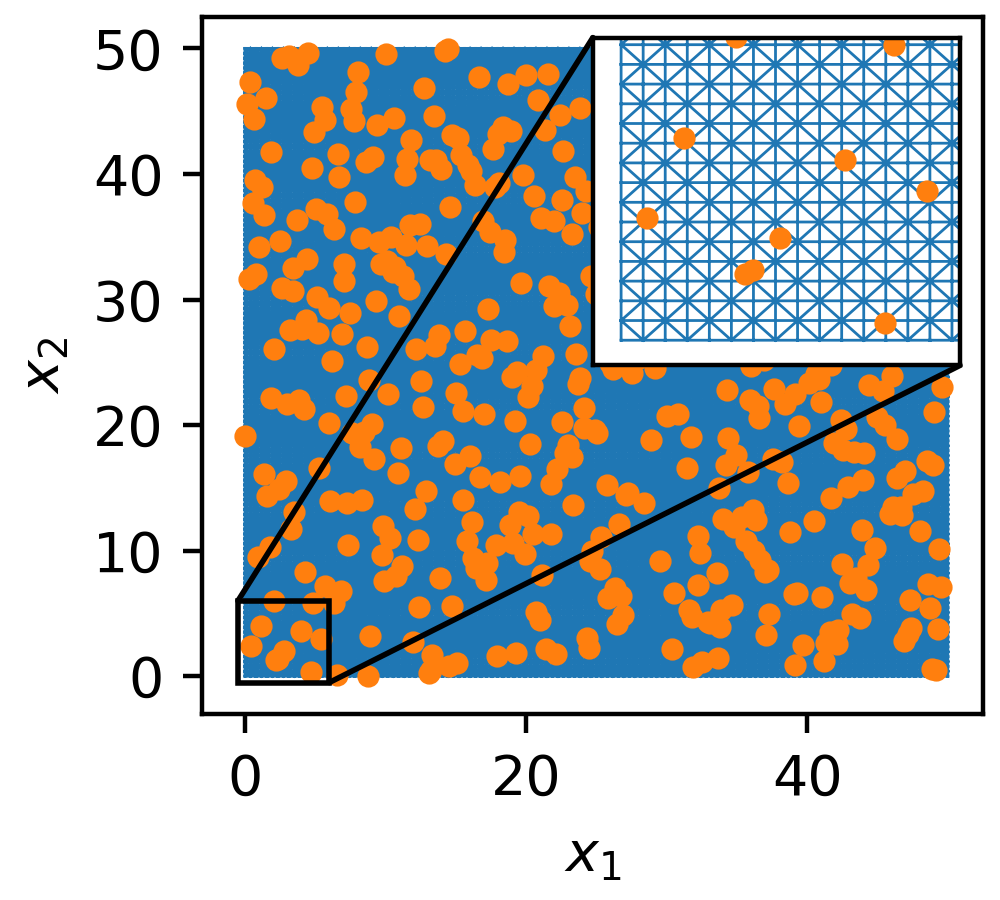}
  \caption{FEM mesh (blue) and observation locations (orange),
    for~\ref{sec:app-parameter-estimation}.}
  \label{fig:stoch-mesh}
\end{figure}

Initial conditions are the same as for the previous oscillatory example,
and the same space-time domain is used
as previous: $\mb{x} \in \Omega = [0, 50] \times [0, 50]$, $t \in [0, 10]$.
Timesteps use an implicit-explicit (IMEX)
scheme~\cite{ascherImplicitExplicitMethodsTimeDependent1995} with forward and
backward Euler:
\[
  \mb{M} \left( \mb{w}_{n + 1} - \mb{w}_n \right)
  + \Dt \kappa \mb{A} \mb{w}_{n + 1}
  = \Dt \tilde{\mb{r}}(\mb{w}_{n}) + \mb{e}_n,
\]
recalling the concatenated state $\mb{w}_n = (\mb{u}_n^\top, \mb{v}_n^\top)^\top$. Timestep
size is set to $\Dt = 10^{-2}$ or $\Dt = 10^{-4}$, and we investigate
the accuracy of the estimated hyperparameters for each (each filter is run for
$1000$ timesteps). Note that this does not yield the same integration window, one
being two orders of magnitude smaller than the other.

\begin{figure}[t]
  \centering
  \includegraphics[width=0.4\linewidth]{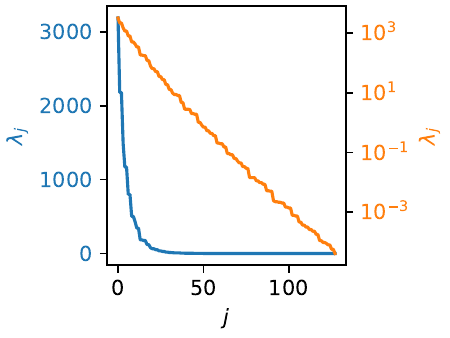}
  \caption{Leading $128$ eigenvalues of $\mb{G}_\theta$
    (for~\ref{sec:app-parameter-estimation}).}
  \label{fig:stoch-G-vals}
\end{figure}

Data, of the $u$ component is observed at $512$ locations at each timestep
(locations shown in Figure~\ref{fig:stoch-mesh}). The assumed data generating
process is $\mb{y}_n = \mb{H} \mb{w}_n + \bm{\eta}_n$,
$\bm{\eta}_n \sim \mathcal{N}(\mb{0}, \sigma^2 \mb{I}_{n_y})$. Data is generated with
$\sigma = 0.01$, and $\bm{\theta} = (\rho, \ell) = (10^{-3}, 10)$.
Filtering is done using LR-ExKF, with $k = k' = 128$ modes, to compute the
posterior $p(\mb{w}_n \given \mb{y}_{1:n}, \bm{\theta}_{1:n}, \sigma_{1:n}, \Lambda) \sim
\mathcal{N}(\mb{m}_n, \mb{L}_n \mb{L}_n^\top)$. For the various filters in this section, each
retain at least $99\%$ of the variance at each timestep. The leading eigenvalues
of $\mb{K}_\theta$, which determine the accuracy of the low-rank approximation
$\mb{G}_\theta^{1/2}$, are shown in Figure~\ref{fig:stoch-G-vals}, and have a range
of approximately $(10^{-4}, 10^3)$.

To avoid recomputing $\mb{G}_\theta$ at each iteration, $\ell$ is fixed at
$\ell = 10$ for all $n$, and we estimate $\rho_n$ and $\sigma_n$ at each $n$,
assuming that $\rho_n$ and $\sigma_n$ are independent for each $n$.
Incorporating more complex temporal structure on these hyperparameters is of
interest and is a possible avenue for future research. Priors are set to the
weakly informative Gaussian priors $\rho_n \sim \mathcal{N}_+(1, 1^2)$ and
$\sigma_n \sim \mathcal{N}_+(0, 1^2)$.

\begin{figure}[t]
  \centering

  \begin{subfigure}[t]{0.45\linewidth}
    \includegraphics[width=\linewidth]{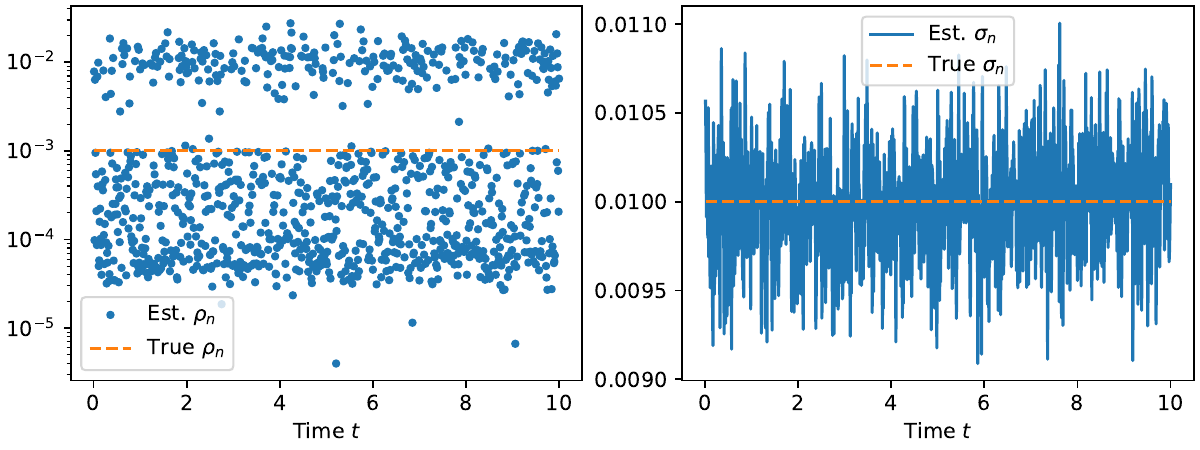}
    \caption{Estimated hyperparameters $\rho_n$ and $\sigma_n$, across all
      times, running the filter with $\Dt = 10^{-2}$.}
    \label{fig:stoch-est-hparam}
  \end{subfigure}
  ~
  \begin{subfigure}[t]{0.45\linewidth}
    \includegraphics[width=\linewidth]{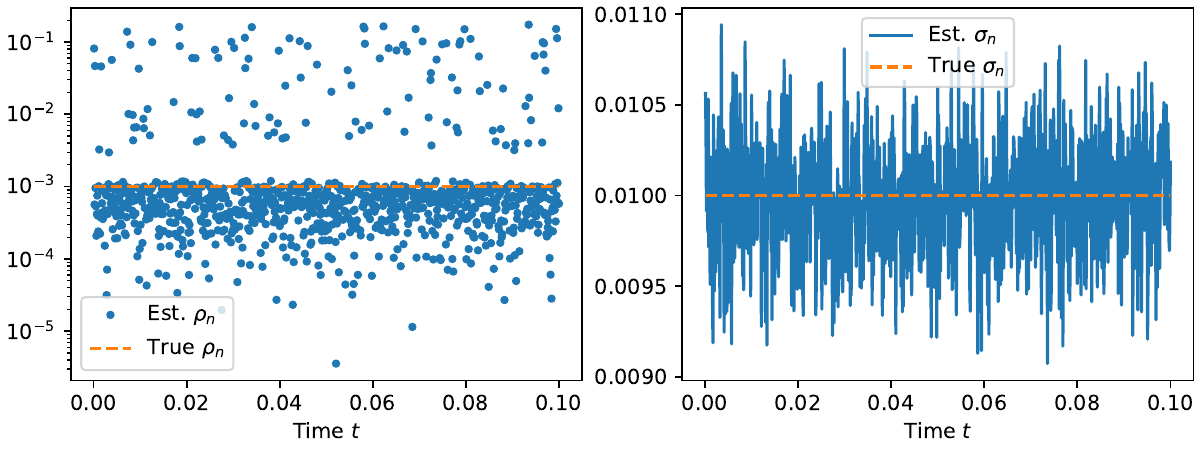}
    \caption{Estimated hyperparameters $\rho_n$ and $\sigma_n$, across all
      times, now running the filter with $\Dt = 10^{-4}$.}
    \label{fig:stoch-est-hparam-small-dt}
  \end{subfigure}

  \begin{subfigure}[t]{0.3\linewidth}
    \includegraphics[width=\linewidth]{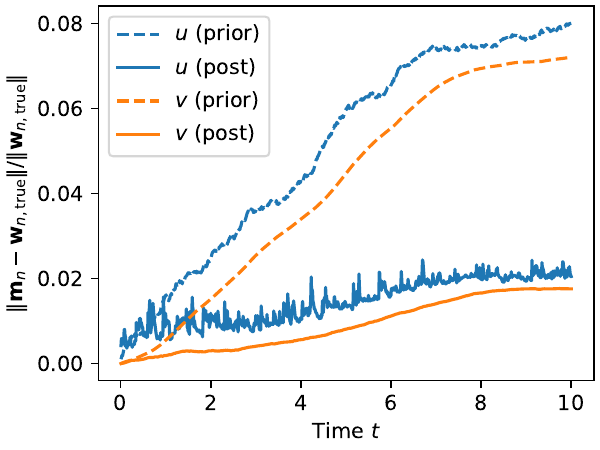}
    \caption{Relative $l^2$ errors on the $u$ (blue) and $v$ (orange)
       components, for the prior (dashed line) and the posterior (solid line),
       using $\Dt = 10^{-2}$.}
    \label{fig:stoch-error}
  \end{subfigure}
  ~
  \begin{subfigure}[t]{0.3\linewidth}
    \includegraphics[width=\linewidth]{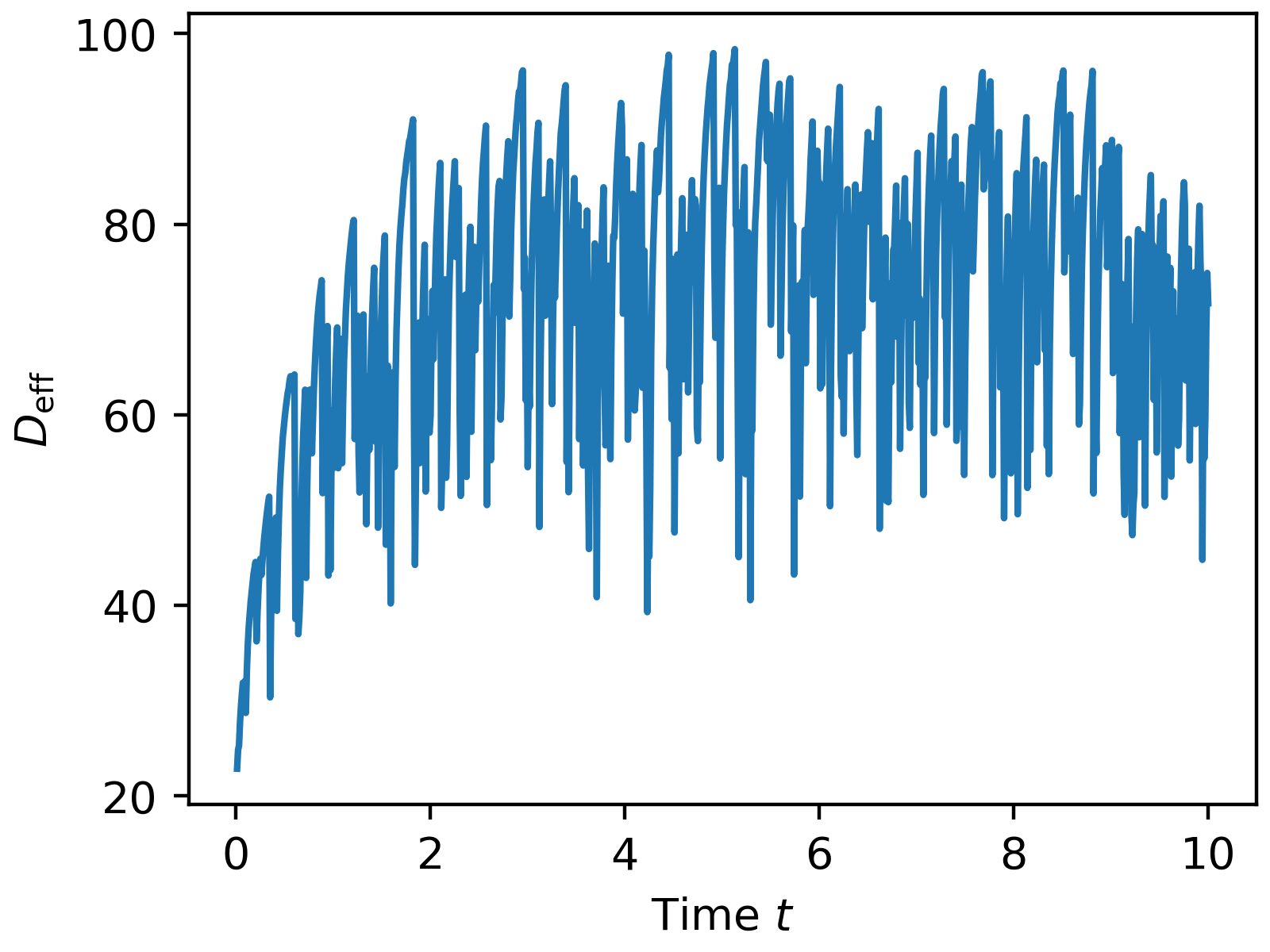}
    \caption{Effective rank $D_{\mathrm{eff}}$ of the prediction covariance
      $\hat{L}_n \hat{L}_n^\top$ using $\Dt = 10^{-2}$.}
    \label{fig:stoch-eff-rank}
  \end{subfigure}
  
  \caption{Diagnostic plots for the stochastic forcing example.}
  \label{fig:stoch-empirical}
\end{figure}

The hyperparameter estimates are shown in
Figures~\ref{fig:stoch-est-hparam}~and~\ref{fig:stoch-est-hparam-small-dt}, and
demonstrate that the noise is identified at each timestep. For the larger
timestep $\Dt = 10^{-2}$ the hyperparameter $\rho_n$ is poorly identified.
Estimates appear to be contained within two point clouds, the topmost cloud
being identified with the noise $\sigma_n$. This inaccuracy is thought to be due
to the combination of linearized dynamics for the covariance update combined
with the low-rank approximation for the square root $\mb{G}_\theta$. Note also
that the inclusion of the truncated Gaussian prior
$\rho_n \sim \mathcal{N}_+(1, 1)$ will also have an effect.

This is confirmed by running the same filter with the smaller timestep $\Dt =
10^{-4}$, thought to increase the accuracy of the linearised covariance
prediction step. The filter appears to better identify the scale
hyperparameter $\rho_n$ (see Figure~\ref{fig:stoch-est-hparam-small-dt}). Some
variation remains, however, which is thought to be due to the low-rank
approximation to $\mb{G}_\theta$ and the truncated Gaussian prior. For completeness
we also plot the relative $l^2$ errors in Figure~\ref{fig:stoch-error} and the
effective rank of the covariance matrix in Figure~\ref{fig:stoch-eff-rank}. We
see that the variation of the effective rank appears to be due to the variation
in the estimates of the hyperparameters (c.f. the smooth effective rank results
seen in the main text for fixed hyperparameter values).

\section{Catastrophic filter divergence in the spiral wave regime}
\label{sec:app-divergence}

\begin{figure}[t]
  \centering
  \includegraphics[width=0.4\linewidth]{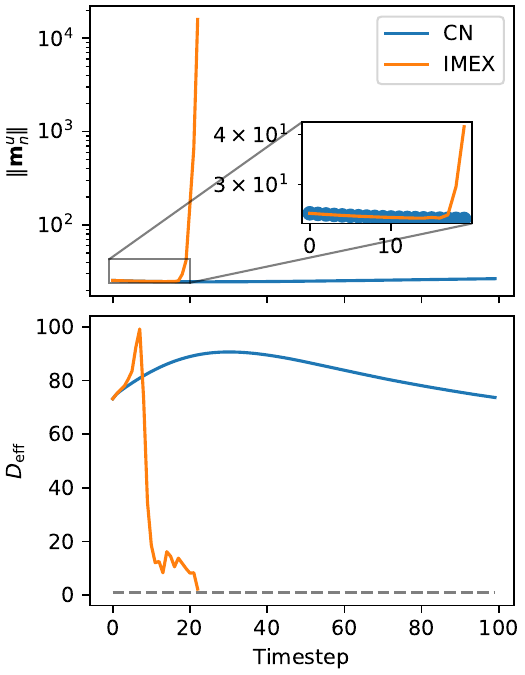}
  \caption{Norm of the statFEM posterior mean $m_n$ for the $u$-component (top), and
    effective rank $D_\mathrm{eff}$ (bottom).}
  \label{fig:spiral-divergence}
\end{figure}

In this case study we show that the LR-ExKF can have catastrophic filter
divergence occur, which is also observed with the
EnKF~\cite{gottwaldMechanismCatastrophicFilter2013,harlimCatastrophicFilterDivergence2010}.
Catastrophic filter divergence is where the posterior mean estimate $\mb{m}_n$
diverges to machine infinity in finite time. Previous studies are contextually
similar, with EnKF divergence occurring in sparsely observed dissipative
nonlinear systems with small noise. The mechanism of the divergence is the use
of an unstable time-integration
scheme~\cite{gottwaldMechanismCatastrophicFilter2013}, which
we verify.

We simulate data according to a deterministic Oregonator in the spiral wave
regime (recall $f = 2$, $q = 0.002$, and $\varepsilon = 0.02$).
These data are observed at every timestep, on the slow $v$ component, across $512$ observation
locations. The initial conditions are the same as those in the second case study
in the main text. Additive Gaussian noise is added, so that $\mb{y}_n = \mb{H}
\mb{w}_n + \bm{\eta}_n$, $\bm{\eta}_n \sim \mathcal{N}(\mb{0}, 10^{-4}
\mb{I}_{n_y})$. There is no model mismatch between the statFEM model and the
underlying data generating process for $\mb{y}_n$; the only difference is that
the statFEM model includes the stochastic process $\xi_v$ on the component $v$,
which has the covariance kernel of Equation~\eqref{eq:gp-forcing}. $\GP$
hyperparameters are fixed, with $\bm{\theta} = (\rho, \ell) = (10^{-3}, 10)$. For the
numerics, both models use an FEM mesh with $128 \times 128$ cells, and the
timestep size is $\Dt = 10^{-3}$.

Filtering is done using the LR-ExKF, with $k = 512$ and $k' = 128$ modes, to compute
the posterior $p(\mb{w}_n \given \mb{y}_{1:n}, \bm{\theta}, \sigma, \Lambda)
\sim \mathcal{N}(\mb{m}_n, \mb{L}_n \mb{L}_n^\top)$. For each
filter more than $99\%$ of the variance is retained at each timestep.. We say that the filter has
diverged if any of elements $\mb{u}_{n, i} \geq 10^4$ for $i = 1, \ldots, n_u$, and two
separate filters are run: one with the IMEX scheme
of~\ref{sec:app-parameter-estimation} for timesteps, and the other with the
Crank-Nicolson (CN) scheme for timesteps.

The IMEX filter diverges to machine infinity after $23$ timesteps (see
Figure~\ref{fig:spiral-divergence}, top). The proposed mechanism of this
divergence is that the posterior estimates of the mean do not accord with the
underlying attractor, resulting in stiffness in the underlying dynamical
model~\cite{gottwaldMechanismCatastrophicFilter2013}. Hence the time integration
becomes stiff, which the IMEX scheme is not able to resolve, and the filter
diverges. In~\cite{gottwaldMechanismCatastrophicFilter2013} this is accompanied
by the effective rank of the covariance matrix reducing to one, which gives
spurious correlations and thus poor posterior estimates in the
update step. This is observed in this scenario, too, with the effective rank
dropping to near unity in finite time by the divergent timestep $n = 23$. Note
also the resemblance to particle filter degeneracy, which results in the
collapse of the particle weights to a Dirac measure.

As in examples in the main text, the behaviour of the relative norm also
``lags'' the behaviour of the effective rank; the effective rank seems to drop
sharply whilst the relative norms of the IMEX and CN are both visually
indistinguishable. Only after the effective rank drops to near unity is the
divergence seen, giving a reminder of the influence of the posterior covariance
matrix $\mb{C}_n$ on posterior estimates of the mean $\mb{m}_n$. Changing the
time integrator to CN results in catastrophic filter divergence being avoided,
and the relative $l^2$ norm and $D_\text{eff}$ appear to asymptotically approach
some limiting value after an initial increase (see
Figure~\ref{fig:spiral-divergence}). Note also the smoothness of the effective
rank $\Deff$ resulting from the fixed choice of hyperparameter $\rho$.

%%% Local Variables:
%%% mode: latex
%%% TeX-master: "rd"
%%% End:

\end{document}